\RequirePackage{tikz}
\RequirePackage[edges]{forest}
\usetikzlibrary{arrows.meta}
\usetikzlibrary{tikzmark}
\usetikzlibrary{bending}
\RequirePackage{multicol}
\usetikzlibrary{shadows}
\RequirePackage{tikz-qtree}

\PassOptionsToPackage{table,prologue}{xcolor}
\documentclass[acmsmall]{acmart}
\AtBeginDocument{%
  \providecommand\BibTeX{{%
    \normalfont B\kern-0.5em{\scshape i\kern-0.25em b}\kern-0.8em\TeX}}}

\usepackage{hyperref}

\usepackage{amsmath,amsfonts}
\usepackage{amsthm}
\usepackage{adjustbox}
\usepackage{tabularx}
\usepackage[vlines]{tabularht}
\usepackage[figuresright]{rotating}
\usepackage{wrapfig}

\usepackage[labelformat=simple]{subcaption}

\usepackage{mathtools}

\usepackage{pifont}
\usepackage{makecell}
\usepackage[T1]{fontenc}

\setcopyright{acmcopyright}
\copyrightyear{2025}
\acmYear{2025}
\acmDOI{XXXXXXX.XXXXXXX}

\acmJournal{CSUR}
\acmVolume{0}
\acmNumber{0}
\acmArticle{000}
\acmMonth{0}

\usepackage{bm}
\newcommand\bolden[1]{{\boldmath\bfseries#1}}

\usepackage{contour}
\usepackage[normalem]{ulem}
\contourlength{0.8pt}

\newcommand{\myuline}[1]{%
  \uline{\phantom{#1}}%
  \llap{\contour{white}{#1}}%
}

\newcommand{\smallsection}[1]{\vspace{0.5mm}{\noindent {\bolden{\myuline{#1}}}}}

\def\mydefbb#1{\expandafter\def\csname bb#1\endcsname{\ensuremath{\mathbb{#1}}}}
\def\mydefallbb#1{\ifx#1\mydefallbb\else\mydefbb#1\expandafter\mydefallbb\fi}
\mydefallbb ABCDEFGHIJKLMNOPQRSTUVWXYZ\mydefallbb

\def\mydefcal#1{\expandafter\def\csname cal#1\endcsname{\ensuremath{\mathcal{#1}}}}
\def\mydefallcal#1{\ifx#1\mydefallcal\else\mydefcal#1\expandafter\mydefallcal\fi}
\mydefallcal ABCDEFGHIJKLMNOPQRSTUVWXYZ\mydefallcal

\newcommand{\abs}[1]{\left\lvert#1\right\rvert}

\newcommand{\setbr}[1]{\{#1\}}
\newcommand{\Set}[1]{\{#1\}}

\definecolor{myred}{HTML}{D62728}
\definecolor{mygreen}{HTML}{2CA02C}
\definecolor{myblue}{HTML}{1F77B4}
\definecolor{mypurple}{HTML}{9467BD}
\definecolor{myorange}{HTML}{FF7F0E}
\definecolor{myolive}{HTML}{BCBD22}
\definecolor{mybrown}{HTML}{8C564B}

\newcommand{\cmark}{\textcolor{mygreen}{\ding{51}}}

\definecolor{peace}{RGB}{228, 26, 28}
\definecolor{love}{RGB}{55, 176, 104}
\definecolor{joy}{RGB}{77, 175, 74}
\definecolor{kindness}{RGB}{152, 78, 163}

\usepackage{xspace}
\newcommand{\hypercl}{\textsc{HyperCL}\xspace}
\newcommand{\hyperlap}{\textsc{HyperLap}\xspace}
\newcommand{\hyperlapp}{\textsc{HyperLap+}\xspace}
\newcommand{\hyperpa}{\textsc{HyperPA}\xspace}
\newcommand{\hyperff}{\textsc{HyperFF}\xspace}
\newcommand{\midas}{\textsc{MiDaS}\xspace}
\newcommand{\thera}{\textsc{THera}\xspace}
\newcommand{\hypertrans}{\textsc{HyperTrans}\xspace}

\let\oldcite\cite
\renewcommand*\cite[1]{\citeauthor{#1}~\oldcite{#1}}

\newcounter{background}
\newcommand{\background}[1]{\refstepcounter{background}\smallsection{B\thebackground. {#1}}}
\newcommand{\backgroundref}[1]{B\ref{#1}\xspace}
\newcommand{\bgref}{\backgroundref}

\newcounter{nullmodel}
\newcommand{\nullmodel}[1]{\refstepcounter{nullmodel}\smallsection{N\thenullmodel. {#1}}}
\newcommand{\nullmodelref}[1]{N\ref{#1}\xspace}
\newcommand{\nmref}{\nullmodelref}

\newcounter{concept}
\newcommand{\concept}[1]{\refstepcounter{concept}\smallsection{E\theconcept. {#1}}}
\newcommand{\conceptref}[1]{E\ref{#1}\xspace}
\newcommand{\ccref}{\conceptref}

\newcounter{measure}
\newcommand{\measure}[1]{\refstepcounter{measure}\smallsection{Q\themeasure. {#1}}}
\newcommand{\measureref}[1]{Q\ref{#1}\xspace}
\newcommand{\msref}{\measureref}

\newcounter{pattern}
\newcommand{\pattern}[1]{\refstepcounter{pattern}\smallsection{P\thepattern. {#1}}}
\newcommand{\patternref}[1]{P\ref{#1}\xspace}
\newcommand{\ptref}{\patternref}

\newcounter{generator}
\newcommand{\generator}[1]{\refstepcounter{generator}\smallsection{G\thegenerator. {#1}}}
\newcommand{\generatorref}[1]{G\ref{#1}\xspace}
\newcommand{\genref}{\generatorref}

\newcounter{dataset}
\makeatletter
\newcommand\footnoteref[1]{\protected@xdef\@thefnmark{\ref{#1}}\@footnotemark}
\makeatother

\usepackage{multirow}

\begin{document}

\title{A Survey on Hypergraph Mining: Patterns, Tools, and Generators}

\author{Geon Lee}
\authornote{Both authors contributed equally to this survey.}
\email{geonlee0325@kaist.ac.kr}
\orcid{0000-0001-6339-9758}
\affiliation{%
  \institution{Kim Jaechul Graduate School of AI, KAIST}
  \city{Seoul}  
  \country{South Korea}  
}
\author{Fanchen Bu}
\authornotemark[1]
\email{boqvezen97@kaist.ac.kr}
\orcid{0000-0003-0497-3902}
\affiliation{%
  \institution{School of Electrical Engineering, KAIST}  
  \city{Daejeon}  
  \country{South Korea}  
}
\author{Tina Eliassi-Rad}
\email{t.eliassirad@northeastern.edu}
\orcid{0000-0002-1892-1188}
\affiliation{%
  \institution{Khoury College of Computer Sciences, Northeastern University}
  \city{Boston}
  \state{MA}
  \country{USA}}
\author{Kijung Shin}
\email{kijungs@kaist.ac.kr}
\orcid{0000-0002-2872-1526}
\affiliation{%
  \institution{Kim Jaechul Graduate School of AI, KAIST}
  \city{Seoul}  
  \country{South Korea}  
}

\renewcommand{\shortauthors}{Lee, Bu, Eliassi-Rad, and Shin}

\begin{abstract}  
Hypergraphs, which belong to the family of higher-order networks, are a natural and powerful choice for modeling group interactions in the real world. For example, when modeling collaboration networks, which may involve not just two but three or more people, the use of hypergraphs allows us to explore beyond pairwise (dyadic) patterns and capture groupwise (polyadic) patterns. The mathematical complexity of hypergraphs offers both opportunities and challenges for hypergraph mining. The goal of hypergraph mining is to find structural properties recurring in real-world hypergraphs across different domains, which we call patterns. To find patterns, we need tools. We divide hypergraph mining tools into three categories: (1) null models (which help test the significance of observed patterns), (2) structural elements (i.e., substructures in a hypergraph such as open and closed triangles), and (3) structural quantities (i.e., numerical tools for computing hypergraph patterns such as transitivity). There are also hypergraph generators, whose objective is to produce synthetic hypergraphs that are a faithful representation of real-world hypergraphs. In this survey, we provide a comprehensive overview of the current landscape of hypergraph mining, covering patterns, tools, and generators. We provide comprehensive taxonomies for each and offer in-depth discussions for future research on hypergraph mining.
\end{abstract}



\received{xx YY 2024}
\received[revised]{xx YY 2024}
\received[accepted]{xx YY 2024}

\maketitle

\section{Introduction}\label{sec:intro}
Group interactions are prevalent in complex real-world systems and appear in various contexts,
including research collaborations~\citep{benson2018simplicial}, 
protein interactions~\citep{feng2021hypergraph}, and 
item co-purchases~\citep{yang2012defining}, to name a few.
These higher-order interactions involving multiple individuals or entities can be naturally and effectively modeled as a hypergraph~\citep{torres2021and,battiston2020networks}.
 
\begin{figure}[t]
	\centering %
	\includegraphics[width=0.9\linewidth]{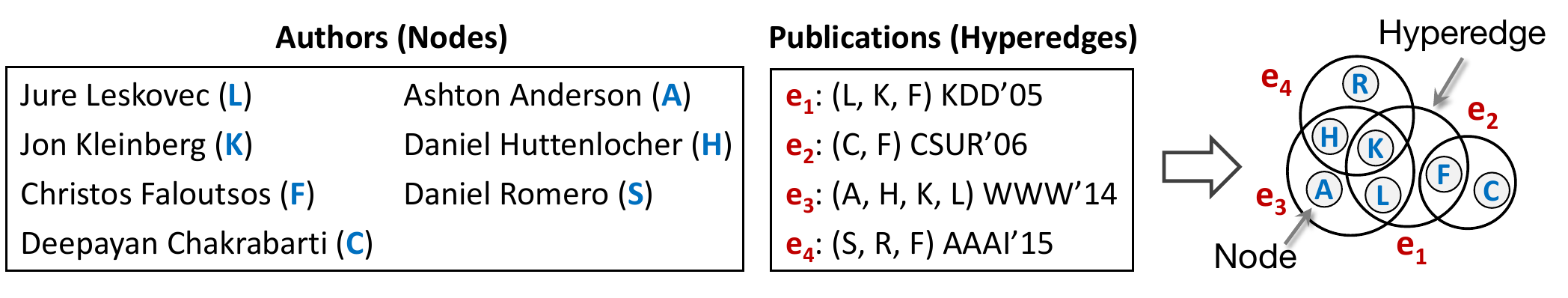}
 \vspace{-3mm}
	\caption{\label{fig:example} 
Group interactions are naturally modeled as hypergraphs. Here, the co-authorship among seven authors across four publications is modeled as a hypergraph with seven nodes and four hyperedges. 
}
\end{figure}

Hypergraphs are a generalization of (pairwise) graphs, consisting of nodes and hyperedges.
Unlike an edge in graphs, which can only connect two nodes, a hyperedge, defined as a non-empty subset of nodes, naturally models an interaction involving any number of nodes.
The flexibility in hyperedge sizes provides hypergraphs with powerful expressiveness, enabling them to accurately model a wide range of group interactions that graphs fall short of. 
For instance, in Figure~\ref{fig:example}, in the co-authorship hypergraph, each node represents a researcher, and each hyperedge represents a co-authorship involving researchers corresponding to its constituent nodes.
It is important to note that co-authorship relationships are not suitably represented by edges in graphs.
When three researchers collaborate on a publication, connecting all possible pairs of researchers fails to distinguish the group interaction from three papers co-authored by different pairs of researchers.
This inherent expressiveness of hypergraphs has led to their applications across a diverse range of fields, including
recommendation systems~\citep{xia2021self},
computer vision~\citep{liao2021hypergraph},
natural language processing~\citep{ding2020more},
social network analysis~\citep{antelmi2021social},
financial analysis~\citep{yi2022structure},
bioinformatics~\citep{feng2021hypergraph},
and 
circuit design~\citep{grzesiak2017hypergraphs}.

Motivated by the successful comprehension of real-world systems using (pairwise) graph modeling (refer to an extensive survey~\citep{chakrabarti2006graph}), recent studies have delved deeply into understanding the structure of real-world hypergraphs, which model these systems.
Hypergraph modeling, particularly the flexibility in the size of each hyperedge, introduces unique perspectives that have not been considered in the context of graphs.
This provides new opportunities and challenges for hypergraph mining, which seeks to enhance our understanding of underlying systems by discovering and explaining the reasons behind structural properties that recur in real-world hypergraphs across different domains.
Such recurring structural properties are often called (structural) patterns.\footnote{For the usage of the concepts of ``hypergraph mining'', ``(structural) patterns'', and ``(mining) tools'', we follow and generalize the convention from the counterparts on pairwise graphs in existing literature~\citep{chakrabarti2022graph,kang2013big,chakrabarti2006graph,jin2020gralsp}.}
Consequently, specialized mining tools (e.g., elements and quantities used to
define structural patterns) have been developed to analyze distinctive structural characteristics of hypergraphs. 
Utilizing these tools, a variety of non-trivial local structural patterns~\citep{lee2020hypergraph,lotito2022higher,benson2018simplicial} and global structural patterns~\citep{kook2020evolution,do2020structural} 
in real-world hypergraphs have been uncovered.
Most of such patterns clearly distinguish real-world hypergraphs from random ones, often accompanied by intuitive explanations or underlying mechanisms. They significantly enhance our understanding of real-world systems.

Hypergraph generators (or hypergraph generative models), are useful for validating our understanding of structural properties.
By reproducing observed patterns, mechanisms in these models offer plausible explanations for real-world patterns.
For this reason, coupled with the prevalence of hypergraph structural patterns,
hypergraph generators have been attracting increasing attention in recent studies~\citep{do2020structural,lee2021hyperedges,benson2018sequences,gallo2023higher}.
These generators successfully produce synthetic hypergraphs that reproduce specific patterns observed in real-world hypergraphs, thereby offering valuable insights for understanding and predicting the structures of hypergraphs.
Such synthetic hypergraphs are also valuable for simulating and evaluating hypergraph algorithms, especially in situations where collecting or tracking real-world hypergraphs is impractical.
Moreover, these generators can be used to create anonymized datasets (specifically, synthetic datasets that are structurally close to a given dataset), as has been widely done on graph data~\citep{murphy2010introducing}.

\begingroup

\setlength{\tabcolsep}{12pt}
\renewcommand{\arraystretch}{0.95}
\begin{table}[t]
    \centering
        \caption{\label{tab:year_range} 
        An overview of the year range for the research works in this survey. 
        `Year' refers to either the publication year or, for unpublished works, the preprint year.
        We only include references where patterns, tools, and/or generators are proposed and/or discussed.}
        \scalebox{0.8}{
        \centerline{
        \begin{tabular}{p{3cm}  p{2cm} p{9.5cm}}
            \hline
		\toprule
            \textbf{Year} & \textbf{\# Papers} & \textbf{References} \\            
            \midrule
            2017 and before & 12 & \citep{cazabet2010detection,estrada2006subgraph,gallagher2013clustering,hein2013total,ihler1993modeling,klamt2009hypergraphs,ramadan2004hypergraph,yoshida2014almost,zhou2006learning,hu2017maintaining,ghoshdastidar2017consistency,li2017inhomogeneous}\\
            \midrule
            2018 & 2 & \citep{benson2018sequences,benson2018simplicial}\\
            \midrule
            2019 & 3 & \citep{ahn2019community,benson2019three,kaminski2019clustering} \\
            \midrule
            2020 & 7 & \citep{chodrow2020configuration,lee2020hypergraph,do2020structural,hayashi2020hypergraphRW,veldt2020local,takai2020hypergraph,kook2020evolution}\\
            \midrule
            2021 & 13 & \citep{lee2021hyperedges,lee2021thyme+,comrie2021hypergraph,limnios2021hcore,tudisco2021node,torres2021and,veldt2021generalized,amburg2021planted,cencetti2021temporal,nakajima2021randomizing,liang2021information,chodrow2021generative,feng2021hypergraph} \\
            \midrule
            2022 & 14 & \citep{lotito2022higher,yuan2022testing,bera2022new,kovalenko2022vector,choo2022persistence,veldt2022hypergraph,Papachristou2022CoreperipheryMF,landry2022hypergraph,pedrood2022hypergraph,kim2022higher,contisciani2022inference,giroire2022preferential,ko2022growth,choe2022midas}\\
            \midrule
            \multirow{2}{*}{2023 and later} & \multirow{2}{*}{28} & \citep{preti2024higher,kim2022reciprocity,lee2023hypergraph,niu2023size,lotito2022exact,moon2023four,lee2023temporal,landry2023simpliciality,arafat2023neighborhood,lee2023k,kim2023exploring,bu2023hypercore,ruggeri2023community,huang2024densest,lanciano2023survey,vasilyeva2023distances,preti2023hyper,Li2023computation,xie2023vital,hu2023identifying,kim2023transitive,ha2023clustering,gallo2023higher,tudisco2022core,kim2023contagion,larock2023encapsulation,choe2024representative,zhang2023efficiently} \\
            \bottomrule
            \hline
        \end{tabular}}     
        }
\end{table}

\endgroup

\smallsection{Scope.}
In this survey, we delve into a broad spectrum of studies on the mining of real-world hypergraphs, aiming to offer a thorough analysis of the current state of the field.
Our survey covers various aspects of hypergraph mining, including 
structural patterns (i.e., structural properties that recur in real-world hypergraphs in different domains),
mining tools (e.g., structural elements and quantities used to define patterns), and
generators that reproduce and thus shed light on patterns.
For each tool, we explain its underlying intuition and its connections to previous concepts.
We provide comprehensive taxonomies (i.e., categorizations) for patterns and generators.
For patterns, we first divide them into static and dynamic ones, based on whether temporal evolution is considered.
We then subdivide the patterns into different levels: node-level, hyperedge-level, subhypergraph-level, and hypergraph-level, according to the minimal element on which each pattern is defined.
For generators, we divide them into full-hypergraph and sub-hypergraph ones, based on whether they generate whole hypergraphs or sub-hypergraphs.
We then subdivide the generators into static and dynamic ones, based on whether they generate static or dynamic (i.e., temporal) hypergraphs.
We systematically compare generators based on their outputs, requirements, and ability to reproduce specific patterns.
To conclude, this survey focuses on patterns that emerge in real-world hypergraphs and generators designed for reproducing these real-world patterns.
Refer to Table~\ref{tab:year_range} for an overview of the year range of the research works covered in this survey.

\smallsection{Related surveys.}
The field of real-world graph mining has a rich historical background, resulting in the development of numerous patterns and generators.
\cite{chakrabarti2006graph} provided a comprehensive overview of patterns in real-world graphs and graph generators.
\cite{drobyshevskiy2019random} and \cite{bonifati2020graph} focused on graph generators, offering a detailed categorization of them.
There has been a rising interest in hypergraphs.
\cite{antelmi2023survey}, \cite{gao2020hypergraph}, and \cite{zhang2022hypergraph} provided a systematic review of hypergraph learning.
\cite{kim2024hgnnsurvey} specifically did an in-depth review of hypergraph neural networks.
\cite{preti2024higher} provided a summary of the advanced analysis techniques for higher-order networks (including hypergraphs).
Some surveys focused on applications of hypergraphs, including visualization~\citep{fischer2021towards} and partitioning~\citep{ccatalyurek2023more}.
\cite{torres2021and} extensively explored different mathematical frameworks, including hypergraphs, for representing higher-order complex systems.
Similarly, \cite{battiston2020networks} examined the usefulness of hypergraphs as a tool for modeling higher-order interactions, from the perspective of dynamical systems and stochastic processes.
There are also several open-source libraries built for hypergraph analysis~\citep{lotito2023hypergraphx,antelmi2019simplehypergraphs,jenkins2018chapel}.
In this survey, we systematically examine structural patterns in real-world hypergraphs, present recent developments in hypergraph mining with a unified taxonomy of hypergraph patterns, and discuss their practical applications for hypergraph generation and other downstream tasks.
Refer to Figure~\ref{fig:roadmap} for a visualized illustration of the structure of this survey.

\begin{figure}[t]
\tikzset{
        basic/.style  = {draw, text width=2cm, align=center, font=\sffamily, rectangle},
        root/.style = {basic, rounded corners=2pt, thin, align=center, fill=green!30, text width=5.5cm},
        onode/.style = {basic, thin, rounded corners=2pt, align=left, text width=4.8cm},
        tnode/.style = {basic, thin, rounded corners=2pt, align=center, fill=pink!60, text width=2cm},
        xnode/.style = {basic, thin, rounded corners=2pt, align=center, fill=blue!20, text width=3.5cm},
        wnode/.style = {basic, thin, align=left, fill=pink!10!blue!80!red!10, text width=6.5em},
    }
    \centering
    \begin{subfigure}[b]{0.45\textwidth}
\centering
\scalebox{0.65}{
    \rotatebox{0}{
    \begin{forest}        
    for tree={
    grow=west,       
    anchor=center,
    parent anchor=west,
    child anchor=east,
  }  
  [
    [\S\ref{sec:intro}: Introduction, xnode, name=nodeIntro, no edge]
    [\S\ref{sec:prelim}: Preliminaries, xnode, name=nodePrelim, no edge, 
    ]
    [\S\ref{sec:tools}: Tools, xnode, name=nodeTools, no edge
        [\S\ref{sec:tools:null_models}: Null models\\
        \S\ref{sec:tools:stru_elements}: Structural elements\\
        \S\ref{sec:tools:stru_qty}: Structural quantities, onode]
    ]
    [\subnode{ptrPatterns}{\S\ref{sec:patterns}: Patterns}, xnode, name=nodePatterns, no edge
        [\S\ref{sec:patterns:static}: Static patterns\\
        \S\ref{sec:patterns:dynamic}: Dynamic patterns, onode]
    ]
  ]  
  \draw[-{Triangle[scale=1]}] (nodeIntro) -- (nodePrelim);
  \draw[-{Triangle[scale=1]}] (nodePrelim) -- (nodeTools);
  \draw[-{Triangle[scale=1]}] (nodeTools) -- (nodePatterns);
  \draw[-{Triangle[scale=1]}] (nodePatterns.east) -- ++(13.8mm,0);;
  \end{forest}
  }
  }
\end{subfigure}
\hfill
\begin{subfigure}[b]{0.54\textwidth}
\centering
\scalebox{0.65}{
    \rotatebox{0}{
    \begin{forest}        
    for tree={
    grow=east,       
    anchor=center,
    parent anchor=east,
    child anchor=west,   
    s sep=0.426cm,
  }
  [
            [\subnode{ptrGen}{\S\ref{sec:generators}: Generators}, xnode, name=nodeGen, no edge
                [\S\ref{sec:generators:full}: Full-hypergraph generators\\
                \S\ref{sec:generators:sub}: Sub-hypergraph generators, onode, fit=band]
            ]
            [\S\ref{sec:future}: Future Applications and Directions, name=nodeFuture, xnode, no edge
                [\S\ref{sec:future:algo_design}: Applications to algorithmic design\\
                \S\ref{sec:future:ml}: Applications to machine learning\\
                \S\ref{sec:future:gen_hgs}: Analysis and mining of generalized hypergraphs, onode, text width=6cm]
            ]
  [\S\ref{sec:conclusion}: Conclusions, name=nodeConc, xnode, no edge]
            ]
  \draw[-{Triangle[scale=1]}] (nodeGen) -- (nodeFuture);
  \draw[-{Triangle[scale=1]}] (nodeFuture) -- (nodeConc);
  \end{forest}
  }
  }
\end{subfigure}
    \caption{The overall structure of this survey.}
    \vspace{-3mm}
    \label{fig:roadmap}
    \color{black}
\end{figure}

\section{Preliminaries}\label{sec:prelim} \label{sec:prelim:background}
In this section, we provide a mathematical background for hypergraphs. %
We use $\bbN \coloneqq \setbr{1, 2, 3, \ldots}$ to denote the set of natural numbers, and we use $[n] \coloneqq \setbr{1, 2, 3, \ldots, n}$ to denote the set of natural numbers at most $n$.
For two sets $A$ and $B$ and a number $k \in \bbN$, we use $A \setminus B \coloneqq \setbr{x \in A: x \notin B}$ to denote the result of set subtraction between $A$ and $B$, and we use $\binom{A}{k} \coloneqq \setbr{A' \subseteq A: \abs{A'} = k}$ to denote the set of all $k$-subsets of $A$.

\background{Hypergraphs.}\label{background:hypergraphs}
A \textit{hypergraph} $H = (V, E)$ is defined by a node set $V$ and a hyperedge set $E$,
where each \textit{hyperedge} $e \in E$ is a subset of $V$, i.e., $e \subseteq V$,
and each \textit{node} $v \in V$ is contained in at least one hyperedge $e \in E$, i.e., $V = \bigcup_{e \in E} e$.
By default, hyperedge weights and repetitions are not considered, and each hyperedge contains at least two nodes (i.e., $\abs{e} \geq 2$).
Explicit clarification will be added for exceptions.
Unlike \textit{(pairwise) graphs} where each edge can only connect two nodes,
each hyperedge in a hypergraph, by definition, can connect an arbitrary number of nodes;
the number $\abs{e}$ of nodes contained in a hyperedge $e$ is called the \textit{size} of $e$.
Given a hypergraph $H = (V, E)$, the \textit{degree} $d(v; H)$ of a node $v \in V$ in $H$ is the number of hyperedges that contain $v$, i.e., $d(v; H) = \abs{\setbr{e \in E: v \in e}}$, and the \textit{volume} $vol(S;H)$ of a subset $S\subseteq V$ of nodes in $H$ is the summation of the degrees of the nodes in the subset, i.e., $vol(S; H)=\sum_{v \in S} d(v;H)$.
\begin{definition}[(Induced) subhypergraphs]\label{def:subhypergraph}
    A hypergraph $H' = (V', E')$ is a \textit{subhypergraph} of another hypergraph $H = (V, E)$, if the hyperedge set of $H'$ is a subset of that of $H$, i.e., $E' \subseteq E$.
    The \textit{induced subhypergraph} of $H$ on a subset $S \subseteq V$ of nodes is $H[S] = (S, \{e \in E : e \subseteq S\}$.
\end{definition}
In some works, the definition of subhypergraphs is more relaxed, and we will provide explicit clarification for such cases.
In this survey, we explicitly use $H$ to denote hypergraphs and use $G$ to imply that we are discussing (pairwise) graphs.

\background{Incidence matrices.}\label{background:incident_matrix}
A straightforward matrix presentation of a hypergraph $H = (V, E)$ is the \textit{incidence matrix} $M_I(H)$ with $\abs{V}$ columns and $\abs{E}$ rows, where each entry represents the membership of a node $v$ in a hyperedge $e$, i.e., $M_I(v, e; H) = 1$ if $v \in e$, and $M_I(v, e; H) = 0$ otherwise.

\background{Paths and connectivity.}\label{background:paths_connectivity}
A \textit{path} consists of a sequence of hyperedges $(e_1, e_2, \ldots, e_{\ell})$ with length $\ell \in \bbN$, where $e_i \cap e_{i + 1} \neq \emptyset, \forall i \in [\ell - 1]$.
A hypergraph $H = (V, E)$ is \textit{connected} if all the node pairs are connected, i.e., for every pair of nodes $v_1, v_2 \in V$, we can find a path $(e_1, e_2, \ldots, e_{\ell})$ with $v_1 \in e_1$ and $v_2 \in e_{\ell}$.
Shortest paths (there can be multiple ones with the same length) between two nodes $v_1$ and $v_2$ are the ones with the smallest length.
A set of hyperedges $E' \subseteq E$ is \textit{connected} if the hypergraph $H' = (V' = \bigcup_{e \in E'} e, E')$ is connected.

\background{Node degree and hyperedge size distributions.}\label{background:degree_size_distribution}
With the basic concepts of node degrees and hyperedge sizes, given a hypergraph $H = (V, E)$, we can numerically summarize $H$ by its node degree and hyperedge size distributions.
Formally, the node degree distribution of $H$ is a function
$\operatorname{dist}_{H}^{(nd)}: \bbN \to [0, 1]$, where
$\operatorname{dist}_{H}^{(nd)}(i) = \abs{\setbr{v \in V: d(v; H) = i}} / \abs{V}$;
and the hyperedge size distribution of $H$ is a function
$\operatorname{dist}_{H}^{(hs)}: \bbN \to [0, 1]$, where
$\operatorname{dist}_{H}^{(hs)}(i) = \abs{\setbr{e \in E: \abs{e} = i}} / \abs{E}$.

\background{Heavy-tailed distributions.}\label{background:heave_tailed_dist}
{Heavy-tailed distributions are defined as distributions with tails heavier than those of exponential distributions~\citep{asmussen2003applied}.
That is, the probability (or frequency) of high values decreases slower in heavy-tailed distributions than in exponential distributions.
As a result, if a quantity follows a heavy-tailed distribution, one may expect to observe that
most values are small, while
extreme high values likely exist.
Two typical types of heavy-tailed distributions are power-law and log-normal distributions.
A widely considered power-law distribution is the Pareto distribution~\citep{arnold2014pareto}.
If a random variable $X_P$ follows a Pareto distribution,
the survival function of $X_P$ has the form 
$\Pr(X_P>x) = \begin{cases}
\left(\frac{x_0}{x}\right)^\alpha & x\ge x_\mathrm{m}, \\
1 & x < x_0,
\end{cases}$,
for some constants $x_m, \alpha > 0$.
A random variable $X_L > 0$ follows a log-normal distribution if and only if 
$\ln(X_L)$ follows a normal distribution.

\begin{figure}[t]
	\centering %
	\includegraphics[width=0.75\linewidth]{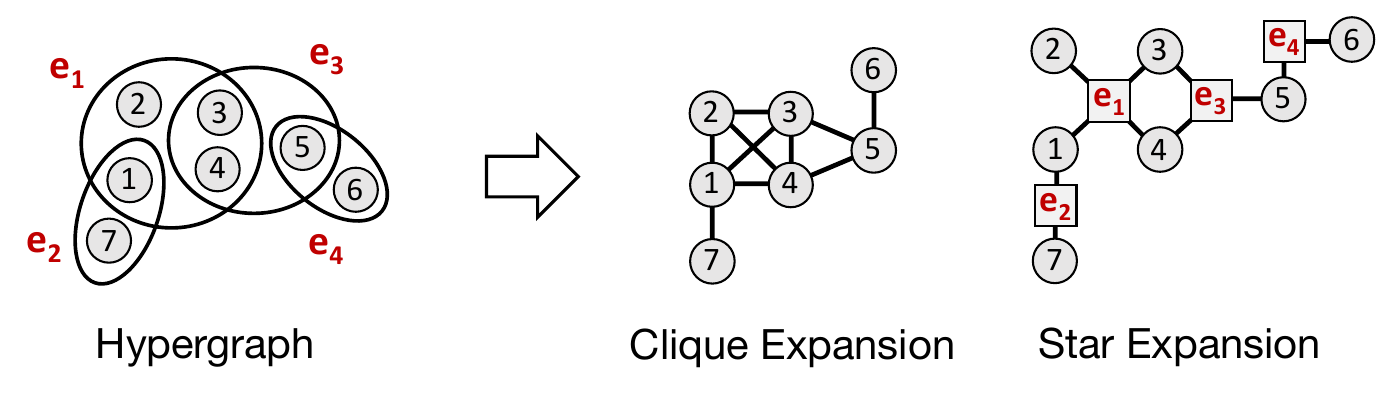}
        \vspace{-2mm}
	\caption{\label{fig:dyadic_projection}         
	Dyadic projections (\bgref{background:dyadic_projections}) are applied to a hypergraph to obtain its clique expansion and star expansion.
}
\end{figure}

\background{Dyadic projections.}\label{background:dyadic_projections}
Dyadic projections approximate hypergraphs by pairwise graphs.
Two common ways of dyadic projections are \textit{clique expansions} and \textit{star expansions}.
{By default, we consider unweighted dyadic projections, and explicit clarification will be added for exceptions.}

\begin{definition}[Clique expansions]
    Given a hypergraph $H = (V, E)$, the \textit{clique expansion} of $H$ is the (pairwise) graph $G_{ce}(H) = (V, E_{ce})$, where $E_{ce} = \bigcup_{e \in E} \binom{e}{2}$.
\end{definition}
Intuitively, clique expansions do not contain higher-order interactions contained in the original hypergraphs,
and information is partially lost.
Refer to a line of research on hypergraph reconstruction (i.e., inference of the missing information) based on pairwise observation~\citep{young2021hypergraph,wang2022supervised}.
One cannot fully recover the original hypergraph from a clique expansion.

\begin{definition}[Star expansions]
    Given a hypergraph $H = (V, E)$, the \textit{star expansion} of $H$ is the (pairwise) graph $G_{se}(H) = (V \cup E, E_{se})$, where $E_{ce} = \setbr{(v, e) \in V \times E: v\in e}$. %
\end{definition}
Although star expansions contain all the incidence information in hypergraphs,
both nodes and hyperedges are uniformly represented as nodes.
However, this symmetrical treatment may not be ideal because of the distinct characteristics of nodes and hyperedges~\citep{yang2022semi}.\footnote{Recently, \citet{yan2024hypergraph} considered a dyadic projection called \textit{cross expansions} by combining clique expansions and star expansions, where this symmetrical treatment is still assumed.}
Nodes and hyperedges are treated distinctly in most hypergraph operations, structural elements, and quantities, thereby breaking this symmetry, which is not achievable in star expansions.
See Figure~\ref{fig:dyadic_projection} for an example of the two dyadic projections introduced above.

\background{Adjacency and Laplacian matrices.}\label{background:adj_laplacian}
With dyadic projections (see \backgroundref{background:dyadic_projections}), one can approximate hypergraphs by pairwise graphs and compute their adjacency and Laplacian matrices.
Given a pairwise graph $G = (V, E)$, the \textit{adjacency matrix}
$M_A(G)$ is with $\abs{V}$ columns and $\abs{V}$ rows, where each entry represents the adjacency between two nodes $v_1$ and $v_2$, i.e., 
$M_A(v_1, v_2; G) = 1$ if $(v_1, v_2) \in E$, and 
$M_A(v_1, v_2; G) = 0$ otherwise.
The \textit{Laplacian matrix} $M_L(G)$ has $\abs{V}$ columns and $\abs{V}$ rows where 
(1) the diagonal is the node degrees (see \backgroundref{background:hypergraphs}), i.e., $M_L(v, v; G) = d(v; G), \forall v$,
and (2) the off-diagonal is the opposite of the adjacency matrix, i.e., $M_L(v_1, v_2; G) = -M_A(v_1, v_2; G), \forall v_1 \neq v_2$. More generalized Laplacian matrices have also been considered for hypergraphs~\citep{saito2018hypergraph,saito2023generalizing}.}
Laplacian matrices are useful for spectral analysis and clustering on graphs~\citep{kang2012heigen} and hypergraphs~\citep{li2018submodular,rodriguez2009laplacian}, closely related to random walks~\citep{Chitra2019RandomWO,carletti2020random,hayashi2020hypergraphRW}, dynamics~\citep{carletti2020dynamical,deArruda2021}, and diffusion processes~\citep{chan2020generalizing,prokopchik2022nonlinear}.
\color{black}

\begin{figure}[t]
	\centering
	\includegraphics[width=0.95\linewidth]{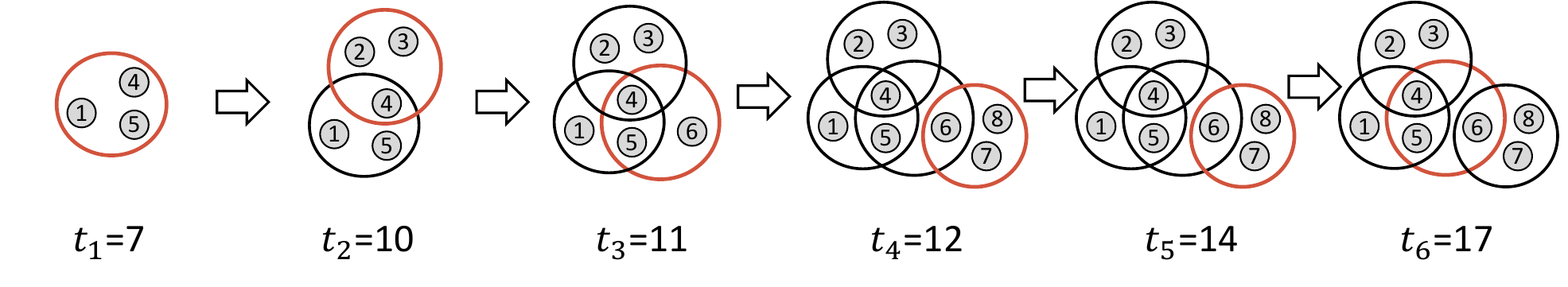}
        \vspace{-2mm}
	\caption{\label{fig:temporal_hypergraph} 
        A temporal hypergraph (\bgref{background:temporal_hypergraphs}) evolves over time as new hyperedges (colored red) are added.
        Note that the same hyperedge can be added multiple times at different time stamps. Here, at $t_4 = 12$ and $t_5 = 14$, the same hyperedge $\{6,7,8\}$ is added.
}
\end{figure}

\background{Temporal hypergraphs.}\label{background:temporal_hypergraphs}
Compared to static hypergraphs introduced above, \textit{temporal hypergraphs} (also called dynamic hypergraphs) describe not only the structural information of hypergraphs but also the temporal evolution.
\begin{definition}[Temporal hypergraphs]
    A \textit{temporal hypergraph} $\calH = (H_0, H_1, \ldots, H_T)$ consists of a series of static hypergraphs called \textit{snapshots} $H_t = (V_t, E_t)$ for each \textit{time step} $0 \leq t \leq T$.
\end{definition}
The temporal information is usually provided as the timestamps of hyperedges.
Suppose a hyperedge $e$ appears $k$ times with timestamps $t_{e0} < t_{e1} < \ldots < t_{ek}$.
By default and in all the works covered in this survey, the hyperedge $e$ is seen as existent after its earliest timestamp $t_{e0}$, i.e.,
$e \in E_{t}, \forall t \geq t_{t0}$.
In such a way, hyperedge deletion is not considered, and explicit clarification will be added for exceptions.
When we consider the temporal structure of hypergraphs, we are able to study some properties that cannot be studied in static hypergraphs.
For example, we can study hyperedge repetitions, i.e., how the same set of nodes appears multiple times at different timestamps.
See Figure~\ref{fig:temporal_hypergraph} for an example temporal hypergraph, where the hypergraph evolves over time as new hyperedges enter it at different time steps.

\section{Tools}\label{sec:tools}

In this section, we introduce mining \textit{tools} for hypergraph structural patterns.\footnote{{See the conventional usage of the term ``tools'' for pairwise graphs in, e.g., \citep{chakrabarti2022graph}.}}
{In general, tools include anything that can be used for defining or mining hypergraph structural patterns.
Typical tools include null models, structural elements, and structural quantities.}
In Figure~\ref{fig:tools:taxonomoy}, we provide an overview taxonomy for the tools we shall introduce below.

\begin{figure}[t]
    \centering
    
    \tikzset{
        basic/.style  = {draw, text width=2cm, align=center, font=\sffamily, rectangle},
        root/.style   = {basic, rounded corners=2pt, thin, align=center, fill=green!30, text width=3.4cm},
        onode/.style = {basic, thin, rounded corners=2pt, align=left, text width=4.8cm},
        tnode/.style = {basic, thin, rounded corners=2pt, align=center, fill=pink!60, text width=2cm},
        xnode/.style = {basic, thin, rounded corners=2pt, align=center, fill=blue!20, text width=2cm},
        wnode/.style = {basic, thin, align=left, fill=pink!10!blue!80!red!10, text width=6.5em},
    }
    \scalebox{0.8}{
    \rotatebox{0}{
    \begin{forest}
        [{Tools}, root,
            [Null Models, xnode,                
                    [
                    \nullmodelref{nullmodel:config_models} Configuration model\\ \nullmodelref{nullmodel:random_filling} Random filling model\\ 
                    \nullmodelref{nullmodel:hypercl} \hypercl, onode]
            ]
            [{Structural Elements}, xnode, text width=3.5cm,             
                    [\conceptref{concept:open_close_triangle} Open and closed triangles\\ 
                    \conceptref{concept:HO-motifs} Higher-order network motifs\\ \conceptref{concept:H-motifs} Hypergraph motifs\\ \conceptref{concept:TH_motifs} Temporal hypergraph motifs\\ \conceptref{concept:ego_networks} Ego-networks\\ \conceptref{concept:multi_lvl_decomp} Multi-level decomposition\\ \conceptref{concept:higher_order_conn} Higher-order connectivity\\ \conceptref{concept:encaps_graph} Encapsulation\\ 
                    \conceptref{concept:hypercore} Hypercores\\                    \conceptref{concept:hypergraph_community} Hypergraph communities\\
                    {\conceptref{concept:dense_substructure} Other dense substrctures}\\
                    , onode]                
            ]
            [{Structural Quantities}, xnode, text width=3.5cm,
                [\measureref{measure:group_deg} Group degrees\\ \measureref{measure:hypercoreness} Hypercoreness\\
                {\measureref{measure:distances} Distances}\\
                {\measureref{measure:centrality_scores} Other centrality scores}\\
                \measureref{measure:hyperedge_homogeneity} Hyperedge homogeneity \\ \measureref{measure:transitivity} Transitivity\\                 
                {\measureref{measure:assortativity} Assortativity} \\
                {\measureref{measure:simpliciality} Simpliciality} \\
                \measureref{measure:char_profile} Characteristic profiles\\ \measureref{measure:density} Density\\ \measureref{measure:overlapness} Overlapness\\ 
                {\measureref{measure:cut} Hypergraph cuts} \\
                {\measureref{measure:conductance} Conductance} \\
                \measureref{measure:modularity} Modularity\\
                \measureref{measure:persistence} Persistence \\
                \measureref{measure:avg_intersection_size} Average intersection size\\
                \measureref{measure:effective_diameter} Effective Diameter
                , onode]
            ]
        ]
    \end{forest}
    }}
    \caption{A taxonomy for tools {used for defining and mining structural patterns} (Section~\ref{sec:tools}).}
    \label{fig:tools:taxonomoy}
\end{figure}

\subsection{Null models}\label{sec:tools:null_models}
We first introduce \textit{null models}.
The concept of null models is very important for significance testing~\citep{raes2007null}, %
where one typically shows that the observed phenomena can hardly happen in null models, and thus shows the observed phenomena are significant, nontrivial, or surprising. %
For pairwise graphs, many random graph models have been used as null models, including
the Erd{\H{o}}s-R{\'e}nyi model~\citep{Erdos1984OnTE} and
the Chung-Lu model~\citep{chung2002average}.
Null models are hypergraph generative models that typically (1) rely on basic information (e.g., node degrees and edge sizes) and (2) lack designs or mechanisms to reproduce realistic patterns beyond the given information.
Thus, they easily fail to capture the properties of real-world hypergraphs in a
comprehensive way.
In contrast, hypergraph generative models  discussed as ``generators'' in Section~\ref{sec:generators}
are designed to effectively reproduce realistic structural patterns.
In addition,  null models and generators serve distinct purposes.
Null models are primarily used for comparison with real hypergraphs, such as in hypothesis testing or validating the significance of a pattern observed in real-world hypergraphs. 
In contrast, generators aim to reproduce realistic patterns observed in real-world hypergraphs, helping to explain and understand the underlying mechanisms that produce these patterns.

\nullmodel{Configuration model.}\label{nullmodel:config_models}
The configuration model is designed to generate random hypergraphs preserving the distributions of node degrees and hyperedge sizes~\citep{chodrow2020configuration}.
This differs from the configuration model for pairwise graphs, which only preserves degree distributions.
Note that there are more advanced hypergraph generators that can be seen as generalized configuration models. We will introduce them in Section~\ref{sec:generators}. 
In practice, one can either use stub matching, which is fast but potentially produces hyperedges with duplicated nodes, 
or use pairwise reshuffling, which avoids hyperedges with duplicated nodes but is slow~\citep{chodrow2020configuration}.

\nullmodel{Random filling model.}\label{nullmodel:random_filling}
The random filling model is a simple variant of the configuration model (see \nmref{nullmodel:config_models})
that preserves hyperedge-size distributions but not node-degree distributions.
Specifically, given a hypergraph, it generates hyperedges with sizes that either precisely follow (or are sampled according to) the original distribution of hyperedge sizes. %
For each hyperedge, the constituent nodes are sampled uniformly at random among all the nodes.

\nullmodel{\hypercl.}\label{nullmodel:hypercl}
The null model \hypercl \citep{lee2021hyperedges} extends the Chung-Lu model~\citep{chung2002average} to hypergraphs.
Given node degrees $\{d_1,\cdots,d_{\vert V\vert}\}$ and hyperedge sizes $\{s_1,\cdots,s_{\vert E\vert}\}$,
\hypercl generates a hypergraph 
with 
$\abs{V}$ nodes $\Set{v_1, v_2, \ldots, v_{\abs{V}}}$ and 
$\abs{E}$ hyperedges $\Set{e_1, e_2, \ldots, e_{\abs{E}}}$ 
in the following manner:
for each hyperedge $e_j$, \hypercl samples $s_j$ nodes with probabilities proportional to the node degrees without replacement,
where the probability of $v_i$ being sampled is proportional to $d_i$.
Hence, the hyperedge sizes are exactly preserved,
and the expected degree of a node $v_i$ is approximately $d_i$.
Empirically, the degree distributions in hypergraphs generated by \hypercl are observed to be close to the input degree distribution~\citep{lee2021hyperedges}, as intended.
{Null models for directed hypergraphs that preserve hyperedge-size and node-degree distributions have also been considered~\citep{preti2024higher}.}

\begin{figure}[t]
\centering 
  \begin{subfigure}[b]{0.2\textwidth}
    \centering
    \includegraphics[width=0.985\columnwidth]{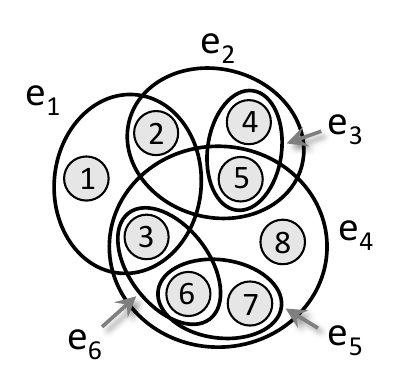}
    \caption{Hypergraph}
    \label{fig:sample_hypergraph}
  \end{subfigure}
  \begin{subfigure}[b]{.2\textwidth}
    \centering
    \includegraphics[width=.45\textwidth]{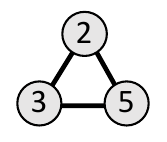}
    \hspace{-3pt}
    \includegraphics[width=.45\textwidth]{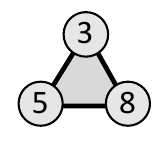}
    \vspace{12pt}
    \caption{Triangles}
    \label{fig:triangle}
  \end{subfigure}
  \begin{subfigure}[b]{.2\textwidth}
    \centering
    \includegraphics[width=.45\textwidth]{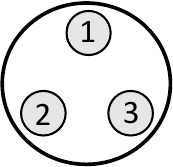}
    \hspace{0.1pt}
    \includegraphics[width=.45\textwidth]{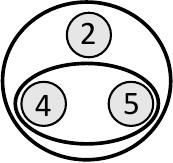}
    \vspace{12pt}
    \caption{HO-motifs}
    \label{fig:ho_motif}
  \end{subfigure}
  \begin{subfigure}[b]{.37\textwidth}
    \centering
    \includegraphics[width=.48\textwidth]{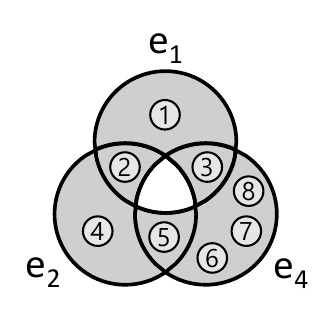}
    \hspace{-3pt}
    \includegraphics[width=.48\textwidth]{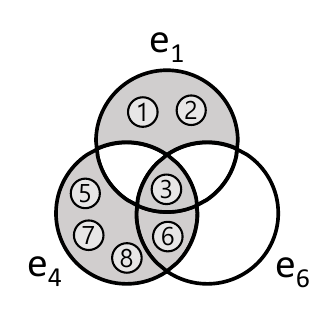}
    \caption{Hypergraph motifs (H-motifs).}
    \label{fig:h_motif}
  \end{subfigure}
\caption{\label{fig:sample}
Different kinds of motifs can be extracted from the same hypergraph.
(a) A hypergraph $H$ with 8 nodes and 6 hyperedges. 
(b; \ccref{concept:open_close_triangle}) Instances of an open (left) and a closed (right) triangle in $H$.  {An edge indicates the existence of a hyperedge containing the two endpoints, and a filled (i.e., closed) triangle indicates the existence of a hyperedge containing all three nodes. 
}
(c; \ccref{concept:HO-motifs}) Two instances of higher-order network motifs (HO-motifs) in $H$.
(d; \ccref{concept:H-motifs}) Two instances of hypergraph motifs (H-motifs) in $H$.}
\end{figure}

\subsection{Structural elements}\label{sec:tools:stru_elements}\label{subsec:concepts}
\textit{Structural elements} include substructures (e.g., subhypergraphs; see Definition~\ref{def:subhypergraph}), as well as the relations and interplays between them. 
These elements help us unravel the underlying structures of hypergraphs and are often objectives upon which structural patterns are defined.

\concept{Open and closed triangles.}\label{concept:open_close_triangle}
Triangles (i.e., three-node cliques) are an important primitive in pairwise graphs as they are used to measure various structural properties, such as community structure~\citep{sotiropoulos2021triangle} and transitivity~\citep{jha2015space}. 
In the context of hypergraphs, triangles can be categorized as \textit{open} and \textit{closed} ones describing different kinds of higher-order interactions among three nodes~\citep{benson2018simplicial}.
As shown in Figure~\ref{fig:triangle}, in an \textit{open triangle}, every pair of nodes co-occurs within one or more hyperedges, but all three nodes do not share any hyperedge.
Conversely, in a \textit{closed triangle}, all three nodes jointly appear in at least one hyperedge.
Notably, this concept can also be extended to higher orders.\footnote{E.g., in an \textit{open quadrangle} involving four nodes, every pair of nodes co-occurs within at least one hyperedge, but all four nodes do not share any hyperedge; while in a \textit{closed quadrangle}, all four nodes jointly appear in at least one hyperedge.}
For instance, consider the hypergraph illustrated in Figure~\ref{fig:sample}.
{Within this hypergraph, nodes $2$, $3$, and $5$ form an open triangle, while nodes $3$, $5$, and $8$ form a closed triangle.}
Importantly, closed triangles, which require at least one hyperedge containing three nodes, cannot be defined in pairwise graphs, and they capture higher-order local structures unique to hypergraphs.

\concept{Higher-order network motifs (HO-motifs).}\label{concept:HO-motifs}
In pairwise graphs, local substructures can be studied by analyzing the occurrences of \textit{network motifs}, which are also known as \textit{graphlets}~\citep{milo2002network,milo2004superfamilies}.
Network motifs, which are classes of structurally identical subgraphs, describe the connectivity patterns among a group of nodes.
\textit{Higher-order network motifs} (HO-motifs) are a natural generalization of network motifs to hypergraphs, and they characterize the local substructures that involve group interactions~\citep{lotito2022higher,lotito2022exact}.
Specifically, 
given a group of $k$ nodes, 
the HO-motifs involving the $k$ nodes are defined by the existence of hyperedges of sizes $2,\ldots,k$ among them.
Note that hyperedges that include nodes other than the $k$ nodes are not considered in the definition.
{For example, in Figure~\ref{fig:ho_motif}, we extract two instances of HO-motifs formed by triples of nodes, $\{1,2,3\}$ and $\{2,4,5\}$, in the hypergraph in Figure~\ref{fig:sample_hypergraph}.
The two instances are distinguished structurally based on the presence of a size-2 hyperedge within a size-3 hyperedge.}
\cite{juul2022hypergraph} proposed a relevant concept where only maximal hyperedges are considered in the motifs.

\concept{Hypergraph motifs (H-motifs).}\label{concept:H-motifs}
While network motifs and higher-order network motifs (see \conceptref{concept:HO-motifs}) describe the connectivity pattern of a fixed number of connected nodes, \textit{hypergraph motifs} (H-motifs) focus on the overlapping patterns of three connected hyperedges~\citep{lee2020hypergraph}.
H-motifs are defined based on the emptiness of each of the seven subsets that correspond to the seven segments in a Venn diagram consisting of three sets.
For three hyperedges $e_i$, $e_j$, and $e_k$, the seven subsets are (1) $e_i \setminus e_j \setminus e_k$, (2) $e_j \setminus e_k \setminus e_i$, (3) $e_k \setminus e_i \setminus e_j$, (4) $e_i \cap e_j \setminus e_k$, (5) $e_j \cap e_k \setminus e_i$, (6) $e_k \cap e_i \setminus e_j$, and (7) $e_i \cap e_j \cap e_k$.
See Figure~\ref{fig:h_motif} for example instances of H-motifs. 
In these instances,
different subsets are non-empty (indicated by shading), and thus their overlapping patterns are distinguished.
Assuming no duplicated hyperedges, a total of 26 unique H-motifs can be defined up to the permutation of hyperedges.
Recently, 
\cite{lee2023hypergraph} and \cite{niu2023size} extended the concept of H-motifs by taking into account not only the emptiness but also the cardinality of the seven subsets. Additionally, \cite{moon2023four} extended this concept to directed hypergraphs \citep{kim2022reciprocity}, where the nodes within each hyperedge are partitioned into head and tail sets, resulting in 91 distinct patterns.

\concept{Temporal hypergraph motifs (TH-motifs).}\label{concept:TH_motifs}
To describe the temporal dynamics of three connected temporal hyperedges, in addition to the overlapping patterns, 96 \textit{temporal hypergraph motifs} (TH-motifs) are defined~\citep{lee2021thyme+,lee2023temporal}. From a structural perspective, TH-motifs follow the concept of H-motifs (see \ccref{concept:H-motifs}) by considering the emptiness of the same seven subsets used in H-motifs. 
In the temporal aspect, TH-motifs are defined for three temporal hyperedges that occur within a short time interval, with a consideration for temporal locality.
In addition, the definition of TH-motifs incorporates the relative arrival order of these three temporal hyperedges, which allows further characterization of patterns that are indistinguishable in the static H-motifs.

\concept{Ego-networks.}\label{concept:ego_networks}
Interactions centering on a single node are commonly analyzed by constructing an \textit{ego-network}~\citep{mcauley2014discovering}, where the center node is called the \textit{ego-node} (or simply \textit{ego}).
An ego-network models the interactions between its ego-node $u$ and the neighbors of $u$ (called the \textit{alter-nodes}, or simply \textit{alters}). %
{\cite{comrie2021hypergraph} defined three types of ego-networks (star ego-network, radial ego-network, and contracted ego-network) by considering different ranges of interactions.}

\begin{figure}[t]
    \centering 
    \includegraphics[width=\columnwidth]{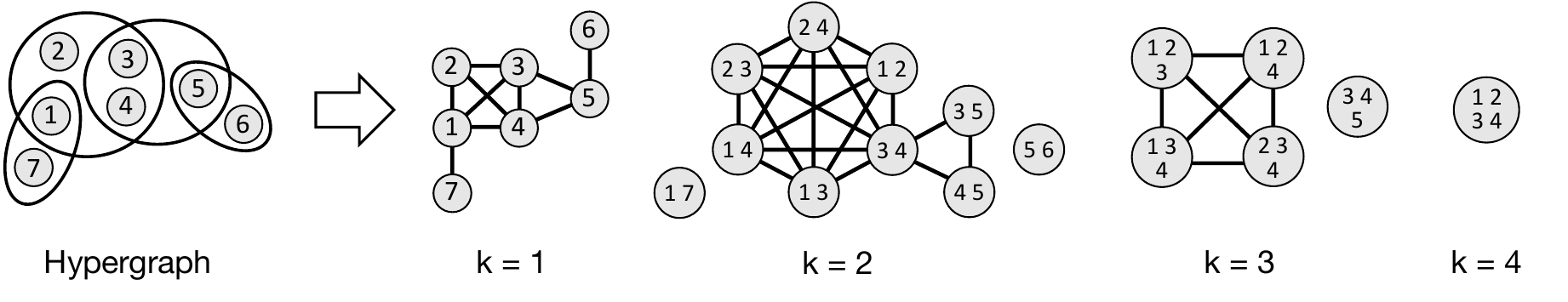}
    \caption{\label{fig:klevel} The multi-level decomposition (\ccref{concept:multi_lvl_decomp}) of a hypergraph. 
    Each $k$-level decomposed graph $G_{(k)}$ models how the node groups of size $k$ interact with other groups.}
\end{figure}

\concept{Multi-Level decomposition.}\label{concept:multi_lvl_decomp}
To reduce the complexity of hypergraphs due to the flexible hyperedge sizes, transforming hypergraphs into pairwise graphs by (connecting) every pair of nodes in each hyperedge (i.e., using clique expansions) has been a common approach. 
However, as discussed in Section~\ref{sec:prelim}, clique expansions may cause a significant amount of information loss, which gives the motivation to develop a more accurate transformation method that preserves the higher-order information.
The \textit{multi-level decomposition}~\citep{do2020structural} of a given hypergraph $H = (V, E)$ gives a series of \textit{$k$-level decomposed graphs}.
In the $k$-level decomposed graph $G_{(k)} =(V_{(k)},E_{(k)})$, each node $v_{(k)}$ represents a \textit{subset} of $k$ nodes that co-exist in at least one hyperedge in $H$, 
and an edge connects two nodes (i.e., subsets) $u_{(k)}$ and $v_{(k)}$ if at least one hyperedge in $H$ contains the union of them.
See Figure~\ref{fig:klevel} for an example of multi-level decomposition.
Intuitively, $k$-level decomposed graphs describe how node groups of size $k$ interact with each other.
Notably, the $1$-level decomposed graph of a hypergraph $H$ is equivalent to the clique expansion of $H$.

\concept{Higher-order connectivity}\label{concept:higher_order_conn}
\cite{kim2022higher} explored an alternative way of transforming hypergraphs into pairwise graphs.
Specifically, given a hypergraph $H = (V, E)$ and a threshold $m$, they proposed to construct a graph $G_{HOC}^{(m)}$ describing the \textit{higher-order connectivity} of $H$, where each hyperedge $e \in E$ is mapped to a node in $G_{HOC}^{(m)}$ and two nodes $e_1 \in E$ and  $e_2 \in E$ are adjacent in $G_{HOC}^{(m)}$ if and only if they share at least $m$ common nodes, i.e.,
$G_{HOC}^{(m)} = (E, E_{HOC}^{(m)})$ with
    $E_{HOC}^{(m)} \coloneqq \Set{(e_1, e_2) \colon e_1, e_2 \in E, \abs{e_1 \cap e_2} \geq m}.$

\concept{Encapsulation.}\label{concept:encaps_graph}
\cite{larock2023encapsulation} proposed the concept of \textit{encapsulation} in hypergraphs.
Specifically, a hyperedge $e_1$ is \textit{encapsulated} by another hyperedge $e_2$ if $e_1 \subseteq e_2$.
{
A \textit{maximal hyperedge} is a hyperedge that is not encapsulated by any other hyperedge~\citep{landry2023simpliciality}.
}

\concept{Hypercores.}\label{concept:hypercore}
The concept of $k$-cores~\citep{seidman1983network}, which is a cohesive subgraph model on pairwise graphs {with various applications (e.g., community detection, %
graph visualization, %
and text analysis; see a survey \citep{malliaros2020core}) %
has been extended to hypergraphs.
We give a collective name, \textit{hypercores}, to such extended concepts.
There are several variants~\citep{arafat2023neighborhood,ramadan2004hypergraph,limnios2021hcore,lee2023k,kim2023exploring,bu2023hypercore},
and we focus on one based on which patterns on real-world hypergraphs are explored.
The concept of $(k, t)$-hypercores~\citep{bu2023hypercore} uses a more general definition of \textit{subhypergraphs} than what we have in Definition~\ref{def:subhypergraph}.
Specifically, the concept of $(k, t)$-hypercores allows not only the hyperedge set to be a subset of the original one (as in Definition~\ref{def:subhypergraph}), but also each hyperedge to be a subset of the original one.
Given a hypergraph $H = (V, E)$, $k \in \bbN$, and $t \in [0, 1]$, the $(k, t)$-hypercore of $H$ is defined as the maximal generalized subhypergraph where 
each node is in at least $k$ hyperedges, and
each hyperedge contains at least $t$ proportion (and at least two) of its original constituent nodes.

\concept{Hypergraph communities.}\label{concept:hypergraph_community}
{The concept of communities (i.e., node groups that are densely connected internally and share relatively sparse connections with outside nodes) has been widely studied on pairwise graphs~\citep{fortunato2010community}. %
The concept has been extended to hypergraphs~\citep{ahn2019community,yuan2022testing,ruggeri2023community}.}
{In hypergraphs, communities are node groups where nodes within the same community are more likely to form hyperedges together, compared to nodes belonging to different communities.
}
{\textit{Clustering} is the process of grouping nodes into communities, and many algorithms have been proposed for discovering communities in hypergraphs~\citep{kaminski2019clustering,feng2023modularity,li2017inhomogeneous}.}

\concept{Other dense substructures.}\label{concept:dense_substructure}
Various definitions of important groups in hypergraphs have been proposed, where a common category defines important groups based on dense substructures.
With the condition of high average degree (i.e., high density; see \measureref{measure:density}) of node subsets, various definitions of ``dense subhypergraphs'' have been proposed~\citep{hu2017maintaining,bera2022new,huang2024densest,lanciano2023survey}.
For example, \cite{musciotto2022identifying} proposed to define dense substructures by how consistently nodes interact.
Recently, \cite{veldt2021generalized} proposed considering the $p$-norm of degree sequence with different $p$ values allowing for flexible emphasis on node degrees, and applied the idea to define generalized dense subgraphs.
The idea can also be generalized to define dense subhypergraphs.

\subsection{Structural quantities}\label{sec:tools-measures}\label{sec:tools:stru_qty}
\textit{Structural quantities} are numerical tools for defining and thus mining hypergraph patterns.
Typically, we compare real-world hypergraphs with random hypergraphs generated by null models w.r.t. specific structural quantities and show the significant numerical differences.

\measure{Group degrees.}\label{measure:group_deg}
As discussed in Section~\ref{sec:prelim:background}, the degree of a node in a hypergraph is defined as the number of hyperedges where the node is contained. 
The concept is naturally extended to groups of nodes and can be employed to study the interrelationships among the appearances of the members of a group.
Specifically, given a hypergraph $H = (V, E)$, the \textit{group degree} $d(S; H)$ of a group of nodes $S \subseteq V$ is defined as the number of hyperedges that 
contain the whole group of nodes, i.e., with a slight abuse of notation,
$d(S; H) \coloneqq \vert \{e\in E : S\subseteq e\} \vert$.

\measure{Hypercoreness.}\label{measure:hypercoreness}
Based on any definition of hypercores (see \ccref{concept:hypercore}), we can have a corresponding measure of \textit{hypercoreness},
{analogous to the concept of coreness in pairwise graphs.}
Since we specifically focus on the concept of $(k, t)$-hypercores~\citep{bu2023hypercore} (see \conceptref{concept:hypercore}), we introduce the corresponding definition of \textit{$t$-hypercoreness} here.
Given a hypergraph $H = (V, E)$ and $t \in [0, 1]$,
the $t$-hypercoreness of a node $v \in V$, denoted by $c_t(v)$, is the maximum $k^*$ such that $v$ is in the $(k^*, t)$-hypercore.
{The hypercoreness of a node essentially quantifies the degree to which each node is centrally positioned within the hypergraph. It has been used for estimating the influence of nodes for contagion models on hypergraphs \citep{bu2023hypercore}
and, together with other quantities, for detecting anomalous nodes \citep{do2023improving}.}

\measure{Distances.}\label{measure:distances}
Based on local connectivity information (e.g., walks and paths; see \backgroundref{background:paths_connectivity}), various distance metrics in hyperedges have been proposed.
\cite{vasilyeva2023distances} and \cite{Li2023computation} proposed distance metrics based on random walks on hypergraphs.
\cite{aksoy2020hypernetwork} proposed a distance metric considering higher-order connectivity in hypergraphs, and \cite{preti2023hyper} proposed a fast approximation algorithm for the metric.
These distance metrics have been used to define node and edge centrality scores~\citep{estrada2006subgraph,yoshida2014almost}, which are applied to real-world tasks, such as critical gene identification~\citep{feng2021hypergraph}.

\measure{Centrality scores.}\label{measure:centrality_scores}
Various scores have been proposed to measure the structural centrality of nodes and edges in hypergraphs. In addition distance-based scores (see~\measureref{measure:distances}), several \textit{spectral centrality scores} have been proposed~\citep{tudisco2021node,kovalenko2022vector,benson2019three}
based on the eigenvalues or eigenvectors of adjacency/incidence/Laplacian matrices (see \backgroundref{background:incident_matrix} and \backgroundref{background:adj_laplacian}).
\cite{xie2023vital} introduced node centrality using a gravity model, while \cite{hu2023identifying} defined centrality based on von Neumann entropy.

\measure{Hyperedge homogeneity.}\label{measure:hyperedge_homogeneity}
The \textit{hyperedge homogeneity} quantifies the degree of structural similarity between the nodes forming a hyperedge together~\citep{lee2021hyperedges}.
Specifically, given a hypergraph $H = (V, E)$, the hyperedge homogeneity of a hyperedge $e\in E$ is defined as:
    $\operatorname{homogeneity}(e; H) \coloneqq \frac{\sum_{S \in {\binom{e}{2}}} d(S;H)}{{\binom{\vert e \vert}{2}}}$.
It corresponds to the average number of hyperedges shared by pairs of nodes within the hyperedge, essentially quantifying the degree of structural similarity among its constituent nodes.
The definition can be extended by considering larger node subsets beyond pairs. %

\measure{Transitivity.}\label{measure:transitivity}
In pairwise graphs, the proportion of two neighbors of a node being adjacent is referred to as \textit{transitivity}~\citep{holland1971transitivity,watts1998collective} or the \textit{clustering coefficient}. %
Several extensions have been proposed for measuring the transitivity of group interactions~\citep{klamt2009hypergraphs,gallagher2013clustering,torres2021and,kim2023transitive}. 
Among them, \hypertrans \citep{kim2023transitive} is a transitivity measure known for its advantageous theoretical properties and its utility in analyzing real-world hypergraphs.
It is defined on \textit{hyperwedges}, i.e., pairs of intersecting hyperedges $e_i, e_j$, and measures the interaction strengths between the two ``wings'' $e_i\setminus e_j$ and $e_j \setminus e_i$.

\measure{Assortativity.}\label{measure:assortativity}
The concept of \textit{assortativity} quantifies the tendency of similar nodes to be adjacent in pairwise graphs~\citep{newman2002assortative}.
A high value of assortativity indicates that similar nodes are more likely to be adjacent compared to dissimilar ones.
The \textit{similarity} is often defined w.r.t. node degrees, i.e., nodes are considered similar if they have similar degrees.
\cite{landry2022hypergraph} extended the concept of assortativity to hypergraphs by capturing how the degree correlation
between nodes within hyperedges deviates from what would be expected in a random one.

\measure{Simpliciality.}\label{measure:simpliciality}
Due to the flexibility in the size of hyperedges, a hyperedge can encapsulate other hyperedges (see \conceptref{concept:encaps_graph}).
The hypergraph simpliciality quantifies how well this hierarchical structure is exhibited in the hypergraph by assessing the extent to which large hyperedges include all possible smaller subsets. 
Specifically, 
the simpliciality of a hypergraph is the ratio of the number of maximal hyperedges that contain all potential subsets to the total number of hyperedges~\citep{landry2023simpliciality}.
A higher ratio indicates that the hypergraph consists of many hyperedges that fully encapsulate all smaller interactions, reflecting a strong inclusion structure.

\measure{Characteristic profiles (CPs).}\label{measure:char_profile}
To better analyze the structural properties of a given hypergraph, we can simultaneously examine multiple structural patterns of interest (e.g., H-motifs) to construct a \textit{characteristic profile} (CP) of the hypergraph.
To do so, the first step is to obtain a numerical frequency of each pattern, e.g., for H-motifs, we can simply count their instances.
Then, for each pattern, by comparing its frequency in the given hypergraph to its frequency in random hypergraphs, we can determine its statistical significance.
Finally, the CP~\citep{milo2004superfamilies,lee2020hypergraph} is a vector that summarizes the structural patterns of the entire hypergraph w.r.t. various patterns, allowing for meaningful comparisons across different hypergraphs that may vary in scale.

\measure{Density.}\label{measure:density}
For pairwise graphs, \textit{density}, defined as the ratio of the edge count to the node count, is a widely-used metric of edge connectivity. Numerous application problems, such as fraud detection in social media, %
expert search in crowdsourcing frameworks, %
and biological module detection %
have been formulated as the task of identifying high-density subgraphs (see a survey \citep{lanciano2023survey}).
The concept of density~\citep{hu2017maintaining,estrada2005complex} naturally extends to (sub)hypergraphs, measuring the ratio of the hyperedge count to the node count in a set $\mathcal{E}$ of hyperedges, i.e., $\rho(\mathcal{E}) \coloneqq \frac{\vert \mathcal{E} \vert}{\vert \bigcup_{e\in \mathcal{E}} e \vert}$.

\measure{Overlapness.}\label{measure:overlapness}
Density (see \msref{measure:density}) provides an intuitive way to measure the degree of overlaps among hyperedges.
However, the effect of hyperedge sizes is overlooked in its definition, leading to quantities sometimes not aligned with intuitive expectations \citep{lee2021hyperedges}.
For instance, consider the
\begin{wrapfigure}{r}{0.64\textwidth}
\centering 
    \includegraphics[width=\linewidth]{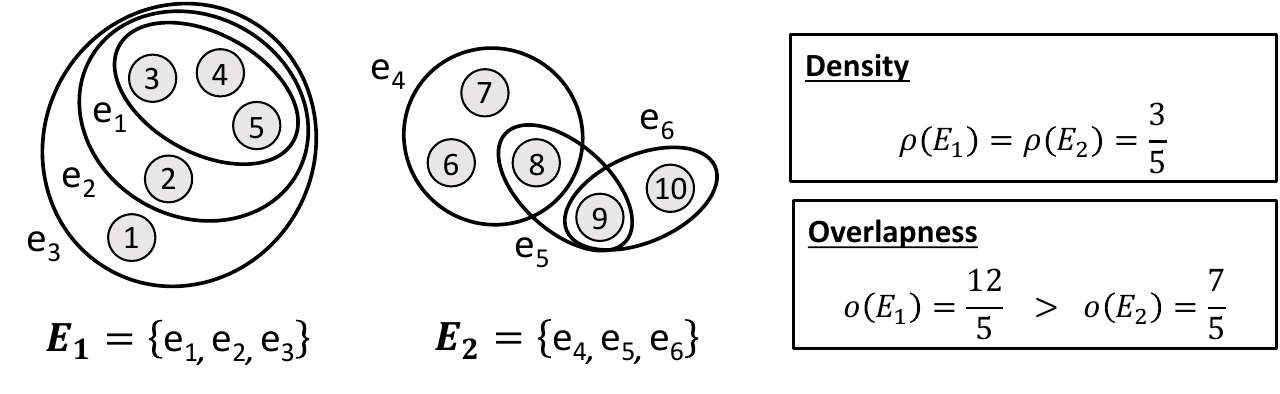}
    \caption{\label{fig:density_overlapness}Two hypergraphs with five nodes and three hyperedges but distinct structures. While they share the same density (\msref{measure:density}), their overlapness  (\msref{measure:overlapness}) differs.}
\end{wrapfigure}
example illustrated in Figure~\ref{fig:density_overlapness}, where the intuitive expectation is that hyperedges in $\mathcal{E}_1$ exhibit more substantial overlaps compared to those in $\mathcal{E}_2$. However, both $\mathcal{E}_1$ and $\calE_2$ have the same density.
\cite{lee2021hyperedges} introduced \textit{overlapness} as an alternative measure that additionally takes the size of individual hyperedges into account.
The overlapness $o(\mathcal{E})$ of a set $\calE$ of hyperedges is 
    $o(\mathcal{E}) \coloneqq \frac{\sum_{e\in \mathcal{E}}\vert e \vert}{\vert \bigcup_{e\in \mathcal{E}} e \vert}$,
equivalent to the average node degree in the sub-hypergraph consisting of the edges in $\calE$.
In line with our intuition,  $\mathcal{E}_1$ in Figure~\ref{fig:density_overlapness} has a higher overlapness than $\mathcal{E}_2$.

\measure{Hypergraph cuts.}\label{measure:cut}
{To assess community strength in hypergraphs (see \ccref{concept:hypergraph_community}), hypergraph cuts generalize the concept of cuts from pairwise graphs~\citep{shi2000normalized}.
The hypergraph cut $cut(S;H)$ for a community $S\subseteq V$ is the sum of \textit{splitting scores} for \textit{boundary hyperedges}~\citep{veldt2022hypergraph,li2017inhomogeneous}, which connect nodes in both $S$ and $V\setminus S$, i.e., $cut(S;H)=\sum_{e\in \partial S} \omega(e,S)$.
The splitting score $\omega(e,S)$ can be defined in various ways~\citep{ihler1993modeling,hayashi2020hypergraphRW,heuer2019network,zhou2006learning,ccatalyurek2023more,deveci2015hypergraph}.
A smaller cut indicates a stronger, more cohesive community.
The community sizes have also been considered when computing hypergraph cuts as normalization terms~\citep{veldt2020local} or constraints~\citep{schlag2023high,gottesburen2024scalable}.}

\measure{Conductance.}\label{measure:conductance}
Conductance evaluates the strength of communities by measuring the ratio between the boundary of a community and the volume (see \backgroundref{background:hypergraphs}) of the community (or the volume of the remaining part).
Given a hypergraph $H=(V,E)$, the conductance $\phi(S;H)$ of a community $S\subseteq V$ is defined as:
    $\phi(S;H) \coloneqq \frac{cut(S;H)}{\min (vol(S;H), vol(V \setminus S; H))}$,
where $cut(S;H)$ is the hypergraph cut (see \measureref{measure:cut}) of the community $S$.
A lower conductance indicates a stronger community.
The conductance~\citep{takai2020hypergraph} of the entire hypergraph $H$ is defined as $\phi(H) = \min_{\emptyset \subsetneq S \subsetneq V} \phi(S;H)$.

\measure{Modularity.}\label{measure:modularity}
{
To assess the strength of community structures (see \ccref{concept:hypergraph_community}) in pairwise graphs, \cite{newman2001clustering} introduced the concept \textit{modularity}.
{A high modularity value implies that node pairs within each community are more likely to have edges between them compared to node pairs belonging to different communities,} and thus it implies strong community structures~\citep{blondel2008fast}.  %
The concept of modularity has been extended to hypergraphs in various ways~\citep{neubauer2009towards,yang2017hypergraph,kaminski2019clustering,giroire2022preferential}.
Modularity measures the difference between the strength of community structures in a given hypergraph and that of a reference random hypergraph.
{Notably, \cite{chodrow2020configuration} considered generalized modularity with various definitions of the reference random hypergraph.}}

\measure{Persistence.}\label{measure:persistence}
{A group of nodes may co-occur in multiple hyperedges over time (e.g., items that are frequently co-purchased). The concept of \textit{persistence} for a group of nodes quantifies how consistently they co-appear over time.}
Formally, given a hypergraph $H = (V, E)$ and a time range $T$, the persistence of a group of nodes $S\subseteq V$ is defined as~\citep{choo2022persistence}:
    $P(S,T; H) \coloneqq \sum_{t \in T} I(S, t; H)$,
where $I(S, t; H) = 1$ if $S$ is a subset of any hyperedge at time $t$, and $I(S, t; H) = 0$ otherwise.
{The persistence of a group can also be understood as its strength or robustness over time.}

\measure{Average intersection size.}\label{measure:avg_intersection_size}
Given a sequence of hyperedges $S_E = (e_1, e_2, \ldots, e_k)$ in chronological order,
\textit{average intersection size} of these hyperedges is defined as \citep{comrie2021hypergraph}: 
    $I(S_E) \coloneqq \frac{\sum_{i = 1}^{k-1} \abs{e_i \cap e_{i + 1}}}{k - 1}$,
That is, this measure calculates the average size of intersections between consecutive hyperedges, which is related to the level of smoothness in temporal changes.

\measure{Effective diameter.}\label{measure:effective_diameter}
To assess the overall connectivity of (hyper)graphs, we can use the concept of \textit{diameter}~\citep{west2001introduction}, which is defined as the longest length of the shortest paths (see \bgref{background:paths_connectivity}) between all the node pairs.\footnote{For each node pair $(v_1, v_2)$, we compute the length of the shortest paths between $v_1$ and $v_2$ and take the largest value.}
Based on that, the concept of \textit{effective diameter} is proposed and used~\citep{leskovec2005graphs}, which is the minimum distance $d$ such that the length of the shortest path(s) between at least $P_{\mathrm{eff}}\%$ (e.g., $P_{\mathrm{eff}} = 90$) node pairs are at most $d$.
{The same definition applies to hypergraphs~\citep{kook2020evolution}.}
The effective diameter is more numerically robust when the shortest path(s) between a few node pairs have extremely high lengths.

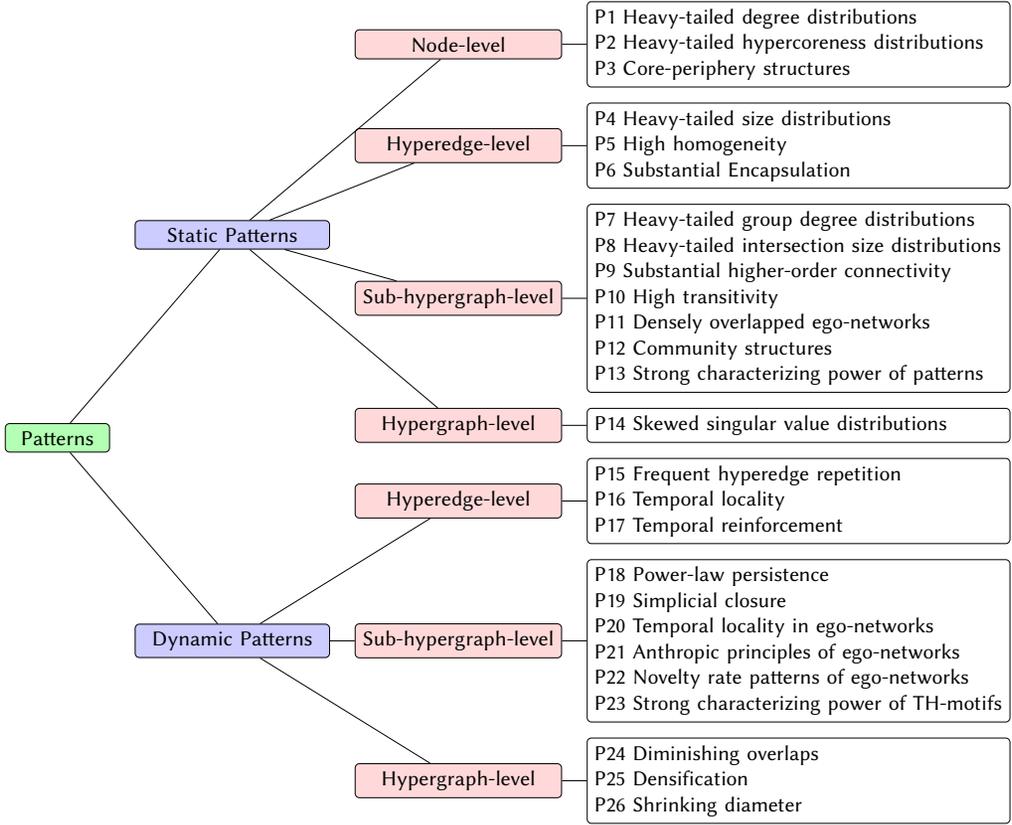
\begin{figure}
    \centering
    \tikzset{
        basic/.style  = {draw, text width=2.8cm, align=center, font=\sffamily, rectangle},
        root/.style   = {basic, rounded corners=2pt, thin, align=center, fill=green!30, text width=1.5cm},
        onode/.style = {basic, thin, rounded corners=2pt, align=left, text width=6.8cm},
        tnode/.style = {basic, thin, rounded corners=2pt, align=center, fill=pink!60, text width=3.2cm},
        xnode/.style = {basic, thin, rounded corners=2pt, align=center, fill=blue!20, text width=3cm},
        wnode/.style = {basic, thin, align=left, fill=pink!10!blue!80!red!10, text width=7cm},
    }
    \scalebox{0.8}{
    \begin{forest}
        for tree={
            l=1.5cm, 
            s sep=0.1cm,
            grow'=0,
            anchor=west,
            child anchor=west,
        }
        [Patterns, root,
            [Static Patterns, xnode,
                [Node-level, tnode,
                    [\patternref{pattern:heavy_tailed_deg_dist} Heavy-tailed degree distributions\\ \patternref{pattern:heavy_tailed_hypercoreness} Heavy-tailed hypercoreness distributions\\ \patternref{pattern:core_periphery} Core-periphery structures, onode]
                ]
                [Hyperedge-level, tnode,
                    [\patternref{pat:heavy_tailed_size_dist} Heavy-tailed size distributions\\ \patternref{pat:hyperedge_homogeneity} High homogeneity\\ \patternref{pat:encapsulation} Substantial Encapsulation, onode]
                ]
                [Sub-hypergraph-level, tnode,
                    [\patternref{pat:heavy_tail_group_deg} Heavy-tailed group degree distributions\\ \patternref{pat:heavy_tailed_intersection_size_distribution} Heavy-tailed intersection size distributions \\ \patternref{pat:higher_order_conn} Substantial higher-order connectivity \\ \patternref{pat:high_transitivity} High transitivity \\ \patternref{pat:density_overlapness} Densely overlapping ego-networks\\ \patternref{pat:community} Community structures\\ 
                    {\patternref{pat:dense_subhypergraph} Dense Subhypergraphs}\\ \patternref{pattern:static:local} Strong characterizing power of patterns, onode]
                ]
                [Hypergraph-level, tnode,
                    [\patternref{pat:singular_value_distribution} Skewed singular value distributions, onode]
                ]            
            ]
            [Dynamic Patterns, xnode,
                [Hyperedge-level, tnode,
                    [\patternref{pat:repeat_behaviors} Frequent hyperedge repetition\\ \patternref{pat:temporal_locality} Temporal locality \\ \patternref{pat:temporal_reinforce} Temporal reinforcement, onode]
                ]
                [Sub-hypergraph-level, tnode,
                    [\patternref{pat:group_persistence} Power-law persistence \\ \patternref{pat:simplicial_closure} Simplicial closure \\ \patternref{pat:egonet_temp_locality} Temporal locality in ego-networks\\ \patternref{pat:egonet_anthropic_principle} Anthropic principles of ego-networks\\ \patternref{pat:egonet_novelty_rate} Novelty rate patterns of ego-networks\\ \patternref{pat:domain_local_patterns_dynamic} Strong characterizing power of TH-motifs, onode]
                ]
                [Hypergraph-level, tnode,
                    [\patternref{pat:diminishing_overlap} Diminishing overlaps\\ \patternref{pat:densification} Densification\\
                    \patternref{pat:diameter} Shrinking diameter, onode]
                ]
            ]
        ]
    \end{forest}}
    \caption{A taxonomy for structural patterns (Section~\ref{sec:patterns}).}
    \label{fig:patterns:taxonomy}
\end{figure}

\section{Structural patterns}\label{sec:patterns}

In this section, we introduce structural \textit{patterns} in real-world hypergraphs {(refer to Table~1 of the supplementary document~\citep{supplementary} for a list of publicly available and frequently used hypergraph datasets from the real world).}
{Structural patterns are structural characteristics that recur in real-world hypergraphs (and the real-world systems they model) across diverse domains~\citep{chakrabarti2006graph}.}
We categorize the structural patterns as follows:
\begin{itemize}
    \item \textbf{Static and dynamic patterns.} Static patterns describe the characteristics of static hypergraphs or individual snapshots of temporal hypergraphs (see \bgref{background:temporal_hypergraphs}), while dynamic patterns describe the evolution of temporal hypergraphs over time.
    Compared to static patterns that focus on structural behaviors, dynamic patterns provide additional insights into temporal behaviors, such as the formation and persistence of group interactions.
    \item \textbf{Node-level, hyperedge-level, subhypergraph-level, and hypergraph-level patterns.} The \textit{level} of a pattern depends on the basic element(s) used to define the pattern. 
    If a pattern describes some properties of individual nodes (or hyperedges), it is categorized as a \textit{node-level} (resp., \textit{hyperedge-level}) pattern.
    Patterns that describe properties of the entire hypergraph are termed \textit{hypergraph-level} patterns.
    Patterns defined on specific combinations of nodes and/or hyperedges are categorized as \textit{subhypergraph-level} patterns.
\end{itemize}
The two forms of categorization are orthogonal, resulting in a total of eight sub-categories formed from their combination.
In Figure~\ref{fig:patterns:taxonomy}, we provide an overview of the taxonomy for the structural patterns that we shall introduce below.
{This section focuses on describing the observed patterns without delving into specific reasons behind them. 
In Section~\ref{sec:generators}, we aim to uncover potential explanations by reproducing them through simple mechanisms. 
}

\subsection{Static patterns}\label{sec:patterns:static}
We first introduce \textit{static patterns}.
These patterns describe the structural properties of nodes and hyperedges, as well as the overall characteristics of real-world hypergraphs. Static patterns do not include those related to temporal changes.

\subsubsection{Node-level patterns}
We shall investigate \textit{static patterns} related to the properties of individual nodes, which are fundamental elements in hypergraphs. %

\pattern{Heavy-tailed degree distributions.}\label{pattern:heavy_tailed_deg_dist}
The degree distributions (see \bgref{background:degree_size_distribution}) of real-world hypergraphs commonly exhibit heavy-tailed distributions~\citep{kook2020evolution}, mostly power-law distributions (see \bgref{background:heave_tailed_dist}).
{This indicates that a small number of nodes are involved in an exceptionally large number of group interactions, while the majority of nodes are involved in only a few interactions.}
Similar patterns have been observed on pairwise graphs,%
\footnote{Pairwise graphs with (asymptotic) power-law degree distributions are called \textit{scale-free} graphs~\citep{barabasi2003scale}.}
and they are (partially) explained by ``the rich get richer''~\citep{faloutsos1999power}, which suggests a temporal process where the nodes with higher degrees are likely to increase their degrees more rapidly as time goes by.
The small number of high-degree nodes are called \textit{hubs}~\citep{barabasi2003scale}, and they play a significant role in many applications. %

\pattern{Heavy-tailed hypercoreness distributions.}\label{pattern:heavy_tailed_hypercoreness}
For a node, its degree can be seen as a centrality measure of it, and so does its hypercoreness (see \msref{measure:hypercoreness}). %
\cite{bu2023hypercore} observed that in real-world hypergraphs, the hypercoreness of the nodes usually exhibits a heavy-tailed distribution (see \bgref{background:heave_tailed_dist}).
{This implies highly dense subhypergraphs involving a small group of nodes, while most nodes do not belong to such subhypergraphs.}
Coreness, the counterpart concept of hypercoreness in pairwise graphs, is also known to commonly exhibit a heavy-tailed distribution in real-world graphs~\citep{shin2016corescope}.
{For many centrality scores (see \measureref{measure:centrality_scores}), although heavy-tailed distributions are not explicitly discussed, there often are nodes with significantly higher centrality values than the other ones~\citep{benson2019three,xie2023vital,hu2023identifying}.}

\pattern{Core-periphery structures.}\label{pattern:core_periphery}
Many real-world hypergraphs have core-periphery structures, where we have 
\textit{core} nodes and \textit{periphery} nodes.
{A large number of hyperedges is supposed to be formed among core nodes,
while periphery nodes are not well-connected to each other (i.e., do not co-exist in many hyperedges) and are supposed to co-exist mainly in hyperedges where at least one core node is present~\citep{tudisco2022core, Papachristou2022CoreperipheryMF,amburg2021planted}.}
Similar structures in pairwise graphs have also been studied~\citep{borgatti2000models}. %

\subsubsection{Hyperedge-level patterns}
We shall now investigate \textit{hyperedge-level static patterns}.
Just like nodes, hyperedges are also fundamental structural elements in hypergraphs (see \bgref{background:hypergraphs}). 
Hence, examining hyperedge-level patterns gives us insights into the structural characteristics of real-world hypergraphs from different perspectives.

\pattern{Heavy-tailed size distributions.}\label{pat:heavy_tailed_size_dist}
A fundamental distinction between pairwise graphs and hypergraphs lies in the fact that hyperedges have variable sizes, connecting any number of nodes, while edges in pairwise graphs can only connect two nodes. Thus, a key property of hyperedges is their size, i.e., the number of nodes that co-occur within a hyperedge.
The distributions of hyperedge sizes in real-world hypergraphs tend to follow heavy-tailed distributions~\citep{kook2020evolution} (see \bgref{background:heave_tailed_dist}).
This implies a large number of small-size hyperedges, while extremely large hyperedges also often exist.

\pattern{High homogeneity.}\label{pat:hyperedge_homogeneity}
The homogeneity (see \msref{measure:hyperedge_homogeneity}) of a hyperedge measures how structurally similar the nodes in the hyperedge are.
\cite{lee2021hyperedges} observed that hyperedges in real-world hypergraphs tend to have significantly higher homogeneity than those in random hypergraphs obtained by the \hypercl model (see \ccref{nullmodel:hypercl}).
This pattern implies that real-world hyperedges are more likely to be filled with structurally similar nodes than randomly chosen nodes.

\pattern{Substantial encapsulation.}\label{pat:encapsulation}
\cite{larock2023encapsulation} studied the encapsulation (see \ccref{concept:encaps_graph}) in real-world hypergraphs by comparing them with random hypergraphs,
where hyperedges of the same size are grouped together and the node labels within each hyperedge group are randomly permuted.
This maintains the overlapping patterns among hyperedges of the same size while randomizing those among hyperedges of different sizes.
They observed that hyperedges in real-world hypergraphs tend to exhibit a significantly higher degree of encapsulation (i.e., encapsulation occurs more frequently) compared to those in corresponding random hypergraphs.
This pattern highlights an aspect of high interconnectedness between hyperedges in real-world hypergraphs.

\subsubsection{Subhypergraph-level patterns}
We shall now investigate \textit{subhypergraph-level static patterns}.
Subhypergraph-level patterns are those regarding neither individual nodes/hyperedges, nor the whole hypergraph.
Instead, they are defined on combinations of nodes and/or hyperedges, e.g., subsets of nodes and hyperedge pairs.

\pattern{Heavy-tailed group degree distribution.}\label{pat:heavy_tail_group_deg}
We have seen the presence of heavy-tailed distributions of (individual) node degrees in real-world hypergraphs (see \ptref{pattern:heavy_tailed_deg_dist}).
{Now we delve into the distributions of degrees of groups of nodes (see \msref{measure:group_deg}) in real-world hypergraphs. }
The distribution of group degrees has been studied by several researchers:
\begin{itemize}
    \item Typically, the group degree of a subset of nodes is the number of hyperedges containing all the nodes in the subset.
    \cite{benson2018sequences} observed that the average degrees of pairs and triples of nodes in real-world hypergraphs are significantly higher than those in random hypergraphs {obtained by the configuration model (see \nmref{nullmodel:config_models})}.
    Moreover, \cite{lee2021hyperedges} observed that the degree distributions of pairs or triples of nodes in real-world hypergraphs {possess thicker tails compared to}
    those in random hypergraphs {obtained by \hypercl (see \nmref{nullmodel:hypercl}).
    This suggests the prevalence of actively co-appearing groups of nodes within real-world hypergraphs}.
    \item \cite{do2020structural} studied group degrees using multi-level decomposition (see \ccref{concept:multi_lvl_decomp}), and they observed that the degree distributions in $k$-level decomposed graphs are heavy-tailed, for different $k$ values.
    Note that the node degrees in a $k$-level decomposed graph are 
    related to but not equivalent to group degrees.%
    
\end{itemize}

\noindent
The above patterns might also be explained by ``the rich get richer'' (see \ptref{pattern:heavy_tailed_deg_dist}; see also \genref{gen:hyperpa}).

\pattern{Heavy-tailed intersection size distributions.}\label{pat:heavy_tailed_intersection_size_distribution}
We have seen that the (individual) hyperedge sizes in real-world hypergraphs follow heavy-tailed distributions (see \ptref{pat:heavy_tailed_size_dist}).
We now extend the scope to hyperedges pairs and study \textit{hyperedge intersections}.
Studying hyperedge intersections allows us to study the connectivity of hypergraphs from a different perspective.
\cite{kook2020evolution} observed that the distributions of intersection sizes of pairs of hyperedges in real-world hypergraphs follow heavy-tailed distributions. 
Moreover, they also observed that some hyperedge pairs in real-world hypergraphs share a substantial number of common nodes (i.e., large intersections), which cannot be observed in random hypergraphs generated by the random filling model 
(see \nmref{nullmodel:random_filling}).

\pattern{Substantial higher-order connectivity.}\label{pat:higher_order_conn}
\cite{kim2022higher} studied hyperedge intersections from another perspective.
They proposed to construct graphs for describing the \textit{higher-order connectivity} of hypergraphs with different thresholds $m$ (see \ccref{concept:higher_order_conn}),
where each hyperedge is represented as a node in the constructed graph, and two nodes are adjacent in the constructed graph if the two corresponding hyperedges share at least $m$ common nodes.
They observed that real-world hypergraphs tend to maintain large connected components with higher values of $m$,
while in random hypergraphs {obtained by the configuration model (see \ccref{nullmodel:config_models})}, indicating substantial hyperedge intersections in real-world hypergraphs, which is in line with the above observations {(see \ptref{pat:heavy_tailed_intersection_size_distribution}).}

\pattern{High transitivity.}\label{pat:high_transitivity}
Several researchers have studied and extended transitivity (see \msref{measure:transitivity}) in hypergraphs, observing high transitivity in real-world hypergraphs from different perspectives:
\begin{itemize}
    \item \cite{kim2023transitive} analyzed the transitivity using the metric \hypertrans, which they proposed (see \msref{measure:transitivity}).
    They observed that, in comparison to random hypergraphs obtained by the \hypercl model (see \ccref{nullmodel:hypercl}), real-world hypergraphs tend to exhibit significantly higher transitivity.
    \item %
    \cite{do2020structural} studied the clustering coefficients in the multi-level decomposed graphs (see \ccref{concept:multi_lvl_decomp}) of real-world hypergraphs.
    They observed that the clustering coefficient of each decomposed graph is significantly higher in 
    real-world hypergraphs than in 
    random hypergraphs generated by the random filling model (see \nmref{nullmodel:random_filling}).
    \item \cite{ha2023clustering} studied local clustering coefficients from the perspective of nodes in real-world hypergraphs.
    {Specifically, they considered \textit{the quad clustering coefficient} of each node $v$, which is defined as the ratio of the actual number of quads incident to $v$ to the maximum possible number of quads incident to $v$, where each quad incident to $v$ is in the form of $(v, v', e_1, e_2)$ with $v \neq v' \in V$, $e_1 \neq e_2 \in E$, and $\{v, v'\} \subseteq e_1 \cap e_2$.} 
    They observed that the average local clustering coefficients in real-world hypergraphs are significantly higher than those in random hypergraphs generated by models similar to the random filling model (see \nmref{nullmodel:random_filling}) or the configuration model (see \nmref{nullmodel:config_models}).
\end{itemize}

\noindent
{These patterns commonly imply that (groups of) nodes are more likely to co-appear in hyperedges
if they share common neighbors.}

\pattern{Densely overlapping ego-networks.}\label{pat:density_overlapness}
The density (see \msref{measure:density}) or overlapness (see \msref{measure:overlapness}) of a hypergraph measures the extent to which its hyperedges overlap with one another. 
\cite{lee2021hyperedges} observed that,
within {star} ego-networks (see \ccref{concept:ego_networks}), the density and overlapness of the star ego-networks in real-world hypergraphs are substantially greater than those in random hypergraphs {obtained by the \hypercl model (see \ccref{nullmodel:hypercl}).} 
This implies that hyperedges in real-world hypergraphs are more locally overlapping than those in random counterparts, which is also related to high transitivity (see \ptref{pat:high_transitivity}).

\pattern{Community structures.}\label{pat:community}
{Communities (see \conceptref{concept:hypergraph_community}) are prevalent in real-world hypergraphs, 
{as demonstrated by the strong community structures quantified
using an extended notion of (normalized) cut (see \measureref{measure:cut})~\citep{veldt2022hypergraph,veldt2020local,li2017inhomogeneous,hayashi2020hypergraphRW}, conductance (see \measureref{measure:conductance})~\citep{hein2013total,veldt2020local,takai2020hypergraph}, and modularity (see \measureref{measure:modularity})~\citep{giroire2022preferential}}.
}
\cite{contisciani2022inference} observed the prevalence of overlapping communities in real-world hypergraphs and proposed a statistical method to detect such communities.
\cite{lotito2023hyperlink} further observed hierarchical {(i.e., communities are further divided into smaller communities)} and multi-scale {(i.e., communities exist at various scales)} {community} structures in real-world hypergraphs.
Notably, \cite{torres2021and} pointed out that different hypergraph representations of the same raw data can display different community structures.

\pattern{Dense subhypergraphs.}\label{pat:dense_subhypergraph}
Real-world hypergraphs often exhibit dense substructures (see \conceptref{concept:dense_substructure}), characterized by high density (see \measureref{measure:density}) or high overlapness (see \measureref{measure:overlapness}).
Their existence is related to the repeated co-occurrence of the same (or similar) set of nodes across multiple hyperedges~\citep{lee2021hyperedges,benson2018sequences}. 
Some studies have shown that real-world hypergraphs tend to exhibit denser subhypergraphs than random hypergraphs~\citep{chodrow2020configuration,lee2021hyperedges}.

\color{black}

\pattern{Strong characterizing power of patterns.}\label{pattern:static:local}
Several measures and patterns can serve as effective characterization tools for hypergraphs.
Specifically, real-world hypergraphs are from diverse domains (or fields), and hypergraphs within the same domain are often observed to be similar w.r.t. some measures and patterns, while ones in different domains are relatively dissimilar.
\begin{itemize}
    \item \cite{benson2018simplicial} used the 
    counts of {open and closed triangles} (see \ccref{concept:open_close_triangle}) to characterize real-world hypergraphs.
    Specifically, the ratio of the number of open triangles and that of closed triangles is a useful metric for distinguishing hypergraphs in different domains.
    \item \cite{lotito2022higher} used the frequencies of {higher-order network motifs} (HO-motifs; see \ccref{concept:HO-motifs}) to characterize real-world hypergraphs.
    Specifically, hypergraphs within the same domain demonstrate similar distributions of the normalized counts of HO-motifs, while hypergraphs from different domains exhibit significant differences.
    \cite{juul2022hypergraph} observed similar phenomena w.r.t. the distribution of {$m$-patterns} (see \ccref{concept:HO-motifs}).
    \item \cite{lee2020hypergraph} used H-motifs (see \ccref{concept:H-motifs}) to construct characteristic profiles (CPs; see \msref{measure:char_profile}) of real-world hypergraphs.
    Such CPs summarize local structural patterns based on H-motifs, and real-world hypergraphs from different domains are clearly distinguished based on CPs.
    \item \cite{larock2023encapsulation} observed that hypergraphs from the same domain exhibit similar encapsulation-related patterns (e.g., how the extent of encapsulation changes w.r.t. hyperedge sizes), while such patterns differ in different domains.
    {
    In addition, \cite{landry2023simpliciality} observed that hypergraphs within the same domain exhibit similar levels of simpliciality (see \measureref{measure:simpliciality}), while hypergraphs from different domains show different levels of simpliciality.
    }
\end{itemize}
The strong characterizing power of structural patterns further validates their usefulness and meaningfulness as tools for analyzing real-world hypergraphs.
Moreover, structural patterns can also be used as structural features \citep{benson2018simplicial,lee2020hypergraph,lee2023hypergraph,lee2021thyme+,lee2023temporal} for machine learning with various downstream applications (see Section~\ref{sec:future:ml}).

\subsubsection{Hypergraph-level patterns}
We shall now investigate \textit{hypergraph-level static patterns}.
Hypergraph-level patterns are regarding the properties of hypergraphs as a whole, and examining them gives us macroscopic insights into real-world hypergraphs.

\pattern{Skewed singular value distributions.}\label{pat:singular_value_distribution}
The singular value decomposition (SVD) of different matrix representations of a real-world hypergraph can provide important insights into the structural properties of the hypergraph.
Specifically, examining the skewness of the singular value distributions can provide information about the underlying hierarchy and community structure (see \ptref{pat:community}) within a hypergraph, as what researchers have done for pairwise graphs~\citep{drineas2004clustering,sarkar2011community,kim2012automated}.
\begin{itemize}
    \item 
    Skewed singular value distributions (i.e., when we sort the singular values in descending order, the values decrease significantly) have been observed for the incidence matrices (see \bgref{background:incident_matrix}) of real-world hypergraphs~\citep{kook2020evolution}.
    \item \cite{do2020structural} considered a different way of matrix representation using multi-level decomposition (see \ccref{concept:multi_lvl_decomp}) and observed skewed singular values of the adjacency matrix of each $k$-level decomposed graph of real-world hypergraphs.    
\end{itemize}

\noindent
The magnitude of singular values reflects the structural importance (e.g., influence) of the corresponding nodes (for incident matrices) or node groups (for the adjacency matrices of multi-level decomposed graphs).
Therefore, the above observations indicate that some nodes or node groups in real-world hypergraphs are considerably more prevalent and influential than others.

\subsection{Dynamic patterns}
\label{subsec:dynamic_patterns}\label{sec:patterns:dynamic}
We now introduce \textit{dynamic patterns}.
Given a temporal hypergraph, dynamic patterns describe the temporal evolution or changes of and within the hypergraph.

\subsubsection{Hyperedge-level patterns}
We shall now investigate \textit{hyperedge-level dynamic patterns}, describing temporal relations between the occurrences of hyperedges.

\pattern{Frequent hyperedge repetitions.}\label{pat:repeat_behaviors}
The recurrence of past events is prevalent in various systems. %
\textit{Hyperedge repetition}, i.e., the recurrence of past hyperedges, has also been observed in the evolution of real-world hypergraphs.
\begin{itemize}
    \item \cite{benson2018sequences} observed that many hyperedges repeatedly appear during the evolution of real-world temporal hypergraphs.
    \item Lee and Shin \citep{lee2021thyme+,lee2023temporal} further observed that the distribution of the number of hyperedge repetitions in real-world temporal hypergraphs typically follows heavy-tailed distributions.
    \item  Lee and Shin \citep{lee2021thyme+,lee2023temporal} also observed that hyperedge repetition happens much more frequently (i.e., the time interval between two occurrences of the same hyperedge is shorter) in real-world hypergraphs than in random hypergraphs.
    {The random hypergraphs were generated by using \hypercl (see \nmref{nullmodel:hypercl}), and the timestamps of the hyperedges are randomly rearranged from the original timestamps.}
    \item \cite{cencetti2021temporal} used the concept of \textit{bursty behaviors} (i.e., the same hyperedge repeats multiple times during a short time interval) and observed that bursty behaviors are more prevalent in real-world hypergraphs than in random hypergraphs obtained by randomly shuffling the timestamps of temporal hyperedges of the same size.
\end{itemize}

\noindent
The authors of the above works commonly observed that hyperedge repetition is much more common in real-world hypergraphs, compared to random counterparts, especially for large hyperedges.

\pattern{Temporal locality.}\label{pat:temporal_locality}
In pairwise-graph evolution, \textit{temporal locality} refers to the tendency for new interactions to be more similar to recent interactions than to older ones~\citep{mahanti2000temporal,lee2020temporal}.
{Examinations of how the structural similarity of hyperedges relates to the timing of their occurrences in real-world hypergraphs
have also revealed the presence of temporal locality.}
\begin{itemize}
    \item \cite{benson2018sequences} observed that newly appeared hyperedges are more likely to share common nodes with hyperedges from the more recent past than with those from the distant past.
    \item For each pair of nodes $v_1$ and $v_2$, and each hyperedge size $k$, \cite{gallo2023higher} counted and compared the numbers of hyperedges of size $k$ containing both $v_1$ and $v_2$ over different time steps.
    They observed temporal correlations, where the aforementioned numbers of the same pair are numerically similar between temporally close time steps.
    {Such temporal correlations exist across different hyperedge sizes.}
    
\end{itemize}
Besides the repeated behaviors of whole hyperedges (see \ptref{pat:repeat_behaviors}), temporal locality, concerning the repetition of subsets within hyperedges, gives unique insights into temporal hypergraph evolution.

\pattern{Temporal reinforcement.}\label{pat:temporal_reinforce}
As a temporal hypergraph evolves, the composition of new hyperedges is influenced by the previous hyperedges.
The phenomenon of \textit{temporal reinforcement} describes how previous hyperedges affect the composition of new hyperedges.
Specifically, \cite{cencetti2021temporal} observed that if a group of nodes has co-appeared in multiple hyperedges in the past, the same group of nodes is more likely to keep co-appearing in some hyperedges in the future, and the likelihood increases as the length of the past co-occurrence period increases.
Such a pattern allows us to analyze the temporal stability of group interactions.

\subsubsection{Subhypergraph-level patterns}
We shall now investigate \textit{subhypergraph-level dynamic patterns}.
Such patterns describe the temporal behaviors of combinations of nodes and/or hyperedges.

\pattern{Power-law persistence.}\label{pat:group_persistence}
In real-world temporal hypergraphs, the same group of nodes may co-occur in multiple hyperedges over time.
The \textit{persistence} of a group of nodes quantifies {how consistently they co-appear over time (see \msref{measure:persistence}).}
\cite{choo2022persistence} observed that in real-world temporal hypergraphs, persistence (specifically, the value of persistence v.s. the number of node groups with such a persistence value) often follows a power-law distribution.
The pattern implies that overall, 
most node groups have a low persistence value, but there are also a small number of node groups that are exceptionally highly persistent.
Such a pattern is related to the above hyperedge-level patterns (see \ptref{pat:repeat_behaviors}-\ptref{pat:temporal_reinforce}).
However, it offers a distinct perspective by taking into account constituent node groups within hyperedges, rather than the entire hyperedges.

\pattern{Simplicial closure.}\label{pat:simplicial_closure}
The concept of \textit{simplicial closure} extends the concept of triadic closure~\citep{simmel1950sociology} in pairwise graphs to hypergraphs, suggesting possible mechanisms by which closed triangles (or higher-order counterparts) are formed (see \ccref{concept:open_close_triangle}).
{\cite{benson2018simplicial} examined the relationship between the emergence of closed triangles (i.e., {the occurrence of a simplicial closure event}) among three nodes in real-world hypergraphs and pairwise connections among them in their clique expansion (see \bgref{background:dyadic_projections}).}
Notably, they considered edge weights by counting edge repetition in clique expansions, i.e., they counted the number of hyperedges each node pair co-appear in.
They observed {the presence of simplicial closure}, i.e., the likelihood of simplicial closure {events} %
tends to increase as the number and/or weights of the connections between the considered nodes increase in the clique expansion.
{This pattern can be readily leveraged for hyperedge prediction  \citep{benson2018simplicial}, extending the utility of triadic closure for link prediction~\citep{huang2015triadic}.}

\pattern{Temporal locality in ego-networks.}\label{pat:egonet_temp_locality}
Identifying the temporal growth patterns of the ego-networks (see~\ccref{concept:ego_networks}) in real-world hypergraphs is an important step toward understanding and predicting the dynamics of group interactions around individual nodes.
Just as the evolution of individual hyperedges exhibits temporal locality (see \ptref{pat:temporal_locality}),
the evolution of ego-networks also shows temporal locality. 
\begin{itemize}
    \item \cite{comrie2021hypergraph} observed that, within an ego-network, hyperedges with closer timestamps also tend to exhibit structural similarity sharing a large number of nodes. 
    Specifically, they measured structural similarity by the average intersection sizes (see \msref{measure:avg_intersection_size}) of temporally consecutive edges in ego-networks.
    Moreover, they observed that this structural similarity decreases as an ego-network evolves and grows over time.
    \item They also explored the temporal locality from the perspective of alter-networks (see \ccref{concept:ego_networks}), and they observed that the average time interval between two consecutive hyperedges within an alter-network is shorter than that in random hypergraphs obtained by randomly shuffling the orders of the hyperedges.   
\end{itemize}
Compared to the temporal locality of hyperedges in the whole hypergraph (\ptref{pat:temporal_locality}), temporal locality within ego-networks provides unique insights into local hypergraph evolution.

\pattern{Anthropic principles of ego-networks.}\label{pat:egonet_anthropic_principle} 
Recall that radial and contracted ego-networks may include hyperedges consisting only of alter-nodes without the ego-node (see~\ccref{concept:ego_networks}).
Consequently, the formation of such ego-networks might have started even before the involvement of the ego-node. 
\cite{comrie2021hypergraph} explored the timing of ego-nodes entering their own ego-networks.  
\begin{itemize}
    \item In contracted ego-networks, we often observe near-perfect positive correlations between the ego-node's arrival time and the size of the ego-network. 
    Specifically, if the ego-network is larger, the ego-node is more likely to arrive later.
    Moreover, ego-nodes tend to arrive {later} in real-world contracted ego-networks than in random ones {obtained by randomly shuffling the orders of the hyperedges}.
    \item {A similar but relatively weaker tendency is frequently noticed in radial ego-networks. 
    In radial ego-networks, even when they are of  considerable size, ego-nodes typically appear before the fifth hyperedge is introduced.}
    Moreover, as in contracted ego-networks, ego-nodes tend to arrive {later} in real-world radial ego networks than in random ones {obtained by randomly shuffling the orders of the hyperedges}.    
\end{itemize}
{These patterns, which give insights into the underlying mechanisms of ego-network construction, are called the \textit{anthropic principles} of ego-networks, with an analogous to how human beings study prehuman histories.}

\pattern{Novelty-rate patterns of ego-networks.}\label{pat:egonet_novelty_rate}
Newly added hyperedges in ego-networks may contain \textit{novel nodes}, which have not previously appeared in the ego-networks. 
The number of such novel nodes is referred to as the \textit{novelty rate}. 
\cite{comrie2021hypergraph} investigated how the novelty rate changes along with ego-network evolution. 
\begin{itemize}
    \item In star and radial ego-networks, the average novelty rate gradually decreases until a certain point. 
    After that point, %
    the novelty rate stays almost constant (for radial ego-networks) or even shows an increasing trend (for star ego-networks).
    \item In contracted ego-networks, the average novelty consistently decreases over time. %
\end{itemize}

\noindent 
Such novel nodes are overall difficult to predict and related to many practical problems, e.g., cold start~\citep{schein2002methods}. 
Understanding the mechanisms behind the appearance of such nodes is both theoretically and practically meaningful.

\pattern{Strong characterizing power of TH-motifs.}\label{pat:domain_local_patterns_dynamic}
Temporal hypergraph motifs (TH-motifs; see \ccref{concept:TH_motifs}) are a temporal extension of the concept of hypergraph motifs (H-motifs) defined in static hypergraphs. 
From the counts of the instances for each of the 96 TH-motif in a given temporal hypergraph, the structural and temporal patterns of the temporal hypergraph can be summarized as a 96-dimensional vector, referred to as the characteristic profile (CP; see \measureref{measure:char_profile}) w.r.t. TH-motifs of the hypergraph.
Utilizing CPs, temporal hypergraphs from different domains can be effectively distinguished, and this distinction is clearer than the differentiation achieved by using H-motifs using only the static information~\citep{lee2021thyme+,lee2023temporal}.
This demonstrates the effectiveness of TH-motifs in characterizing temporal hypergraphs by capturing both temporal and structural patterns.

\subsubsection{Hypergraph-level patterns}
We shall now investigate \textit{hypergraph-level dynamic patterns}.
These patterns describe how the characteristics of hypergraphs as a whole change over time.

\pattern{Diminishing overlaps.}\label{pat:diminishing_overlap}
\cite{kook2020evolution} investigated how the structural interconnectedness of hyperedges evolves over time in real-world temporal hypergraphs. 
Specifically, they observed that the proportion of intersecting hyperedge pairs among all hyperedge pairs tends to decrease over time.
{This observation aligns with the finding that the similarity between hyperedges diminishes as the time gap between their occurrences increases (see \ptref{pat:temporal_locality}).}

\pattern{Densification.}\label{pat:densification}
In pairwise graphs, densification is characterized by the phenomenon where the density (see \msref{measure:density}) of graphs increases over time, observed by~\cite{leskovec2007graph}.
Specifically, they observed that the number of edges increases superlinearly w.r.t. the number of nodes (i.e., $\abs{E} \propto \abs{V}^{1 + \alpha}$ with $\alpha > 0$), and thus
the density $\frac{\abs{E}}{\abs{V}} \propto \abs{V}^{\alpha}$ increases as graphs grow.
\cite{kook2020evolution} observed similar trends in the evolution of real-world hypergraphs, 
where the number of hyperedges increases superlinearly w.r.t. the number of nodes, and thus the density of real-world hypergraphs also increases over time as they grow.

\pattern{Shrinking diameter.}\label{pat:diameter}
Shrinking diameter is another pattern observed in real-world pairwise graphs~\citep{leskovec2007graph}, where the effective diameter (see \msref{measure:effective_diameter}) often decreases as graphs grow.
Such trends can be naturally extended to hypergraphs, and
\cite{kook2020evolution} indeed observed such trends in the evolution of real-world hypergraphs.
This suggests that information or influence may spread more rapidly as real-world hypergraphs expand in size.

\section{Generators}\label{sec:generators}

\begin{figure}
    \centering
    
    \tikzset{
        basic/.style  = {draw, text width=2cm, align=center, font=\sffamily, rectangle},
        root/.style   = {basic, rounded corners=2pt, thin, align=center, fill=green!30},
        onode/.style = {basic, thin, rounded corners=2pt, align=left, text width=2.8cm},
        tnode/.style = {basic, thin, rounded corners=2pt, align=center, fill=pink!60, text width=2cm},
        xnode/.style = {basic, thin, rounded corners=2pt, align=center, fill=blue!20, text width=2.5cm},
        wnode/.style = {basic, thin, align=left, fill=pink!10!blue!80!red!10, text width=6.5em},
        edge from parent/.style={draw=black, edge from parent fork right}
    }
    \scalebox{0.82}{
    \rotatebox{0}{
    \begin{forest}
        [Generators, root,
            [Full-hypergraph, xnode,
                [Static, tnode,
                    [                    
                    \generatorref{gen:hyperlap} \hyperlap \;\;\;\hspace{0.4pt}
                    \generatorref{gen:cigam} CIGAM \\
                    \generatorref{gen:hyper_dk} \textsc{Hyper}-$dK$ \;\;\;    {\generatorref{gen:assortative} AMHM}\\              {\generatorref{gen:hyperSBM} HSBM} \;\;\;\;\;\;\;\; \generatorref{gen:rnhm} RNHM  \\ \generatorref{gen:hoc} HOC \\
                    ,onode, text width=4.2cm
                    ] 
                ]
                [Dynamic, tnode,
                    [
                    \generatorref{gen:hyperpa} \hyperpa \;\;\;\;
                    \generatorref{gen:hmpa} HMPA  \\
                    \generatorref{gen:hyperff} \hyperff \;\;\hspace{1pt}
                    \generatorref{gen:thera} \thera \\
                    \generatorref{gen:darh} DARH  \;\;\;\;\;\;
                    ,onode, text width=4.3cm]
                ]
            ]
            [Sub-hypergraph, xnode,
                [Static, tnode,
                    [\generatorref{gen:midas} \midas \\
                    \generatorref{gen:hrw} HRW, 
                    onode, text width=2cm, text height=0.2cm]
                ]
                [Dynamic, tnode,
                    [\generatorref{gen:trhc} TRHC \\ \generatorref{gen:cru} CRU,  onode, text width=2cm]
                ]
            ]
        ]
    \end{forest}
    }}
    \caption{Taxonomy for generators (Section~\ref{sec:generators}).}
    \label{fig:generators:taxonomoy}
\end{figure}
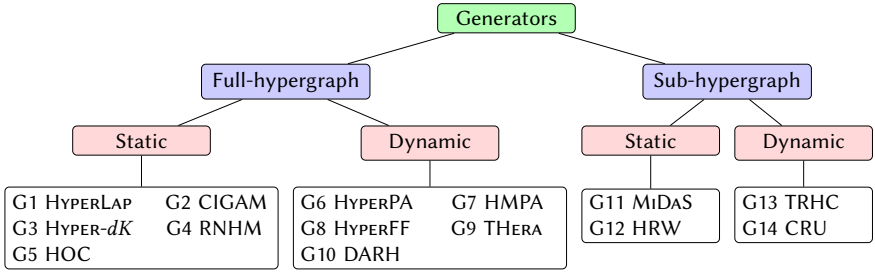

In this section, we introduce hypergraph \textit{generators}.
We focus on the generators based on the properties of real-world hypergraphs.
Hypergraph generators play a crucial role in large-scale modeling by enabling the creation of synthetic datasets that mimic real-world hypergraph structures, facilitating benchmarking and scalability testing.
In Figure~\ref{fig:generators:taxonomoy}, we provide the taxonomy for the generators that we shall discuss below.
We categorize {the generators of our interest} into:
\begin{itemize}
    \item \textbf{Full-hypergraph and sub-hypergraph generators.} 
    Full-hypergraph generators generate whole hypergraphs, while sub-hypergraph generators generate parts of hypergraphs.
    \item \textbf{Static and dynamic generators.} 
    {Static generators generate static hypergraphs,
    while dynamic generators generate dynamic graphs (i.e., temporal hypergraphs; see \bgref{background:temporal_hypergraphs}).
    Notably, in some works, the authors do not explicitly mention whether the proposed generator generates static hypergraphs or dynamic hypergraphs.
    For such cases, we categorize generators based on whether the generation process can be interpreted as an evolving process with temporal dependency.}
    {For example, generators based on preferential attachment can be naturally interpreted as evolving processes, while generators using node shuffling or rewiring cannot.}
\end{itemize}
The two kinds of categorization are orthogonal, and thus we have in total four sub-categories. 

For each generator, we provide a summary of its algorithmic process with intuitions and discussions, if any.
See Table~\ref{tab:generators:input_output} for the detailed input and output of each generator.

\subsection{Full-hypergraph generators}\label{sec:generators:full}
We first introduce \textit{full-hypergraph generators}.
A full-hypergraph generator takes some hypergraph statistics as the input (usually with some additional model-specific hyperparameters) 
and outputs a \textit{whole} hypergraph that is supposed to preserve some structural patterns in real-world hypergraphs.

\subsubsection{Static generators}
We shall introduce \textit{static full-hypergraph generators} below.

\begingroup
\renewcommand{\arraystretch}{0.9}
\begin{table}[t!]
\centering
\caption{\label{tab:generators:input_output}We compare hypergraph generators based on their input information: the number of nodes (NN), a node degree distribution (DD), a hyperedge size distribution (SD), other information about input hypergraph (OI), an entire hypergraph (EH), and the number of hyperparameters (HP).
We also list each generator's output information, which includes the type of output (same as the categorization of the generator) and the patterns validated (i.e., reproduced) by the generator.}
\scalebox{0.72}{
\centerline{
\begin{tabular}{lcccccccl}
    \toprule
     & \multicolumn{6}{c}{\textbf{Input Information}} & \multicolumn{2}{c}{\textbf{Output Information}} \\
     \cmidrule(lr){2-7} \cmidrule(lr){8-9}
     \multicolumn{1}{l}{\textbf{Generator}} & \textbf{NN} & \textbf{DD}* & \textbf{SD}* & \textbf{OI} & \textbf{EH}* & \textbf{HP} & \textbf{Type/Categorization} & \textbf{Validated Patterns*}
     \\ \midrule
    {Configuration model (\nmref{nullmodel:config_models})} 
    & \cmark & \cmark & \cmark &  &  
    & - & 
    Static Full-Hypergraph & 
    \ptref{pattern:heavy_tailed_deg_dist}, \ptref{pat:heavy_tailed_size_dist}
    \\      
    {Random filling model (\nmref{nullmodel:random_filling})} 
    & \cmark & & \cmark &  &  
    & - & 
    Static Full-Hypergraph &     
    \ptref{pat:heavy_tailed_size_dist}
    \\ 
    {\hypercl (\nmref{nullmodel:hypercl})} 
    & \cmark & \cmark & \cmark &  &  
    & - & 
    Static Full-Hypergraph & 
    \ptref{pattern:heavy_tailed_deg_dist}, \ptref{pat:heavy_tailed_size_dist}
    \\     
    \midrule
    {\hyperlap (\genref{gen:hyperlap})} 
    & \cmark & \cmark & \cmark & \cmark &  
    & $O(\log \abs{V})$** & 
    Static Full-Hypergraph & 
    
    \ptref{pattern:heavy_tailed_deg_dist}, 
    \ptref{pat:heavy_tailed_size_dist}, 
    \ptref{pat:hyperedge_homogeneity},
    \ptref{pat:heavy_tail_group_deg},
    \ptref{pat:density_overlapness}
    \\
    CIGAM (\genref{gen:cigam}) 
    & \cmark & & & & 
    & $O(\abs{V})$ &
    Static Full-Hypergraph & 
    \ptref{pattern:heavy_tailed_deg_dist}, 
    \ptref{pattern:core_periphery} 
    \\
    \textsc{Hyper}-$dK$ (\genref{gen:hyper_dk})
    & \cmark & \cmark & \cmark & \cmark &  
    & $O(1)$ & 
    Static Full-Hypergraph & 
    \ptref{pattern:heavy_tailed_deg_dist}, \ptref{pat:heavy_tailed_size_dist}, 
    \ptref{pat:heavy_tail_group_deg}
    \\ 
    {AMHM (\genref{gen:assortative})}
    & \cmark & \cmark & \cmark & \cmark &  
    & {$O(\abs{E})$} &
    {Static Full-Hypergraph} & 
    {\ptref{pattern:heavy_tailed_deg_dist}, \ptref{pat:heavy_tailed_size_dist}}
    \\
    {HSBM (\genref{gen:hyperSBM})}
    &  &  &  &  & \cmark
    & {$O(\abs{E})$} &
    {Static Full-Hypergraph} & 
    {\ptref{pattern:heavy_tailed_deg_dist}, \ptref{pat:heavy_tailed_size_dist}, \ptref{pat:community}}
    \\    
    RNHM (\genref{gen:rnhm})
    & \cmark &  &  &  &  
    & $O(\abs{V})$ &
    Static Full-Hypergraph &
    \ptref{pat:encapsulation}
    \\
    HOC (\genref{gen:hoc})
    & \cmark &  & \cmark &  &  
    & $O(\abs{V})$ &
    Static Full-Hypergraph &
    \ptref{pat:heavy_tailed_size_dist},
    \patternref{pat:higher_order_conn}
    \\
    \midrule
    \hyperpa (\genref{gen:hyperpa}) 
    & \cmark &  & \cmark & \cmark &  
    & - & 
    Dynamic Full-Hypergraph &
    \ptref{pattern:heavy_tailed_deg_dist}, 
    \ptref{pat:heavy_tailed_size_dist},
    \patternref{pat:heavy_tail_group_deg}, \patternref{pat:high_transitivity},
    \patternref{pat:singular_value_distribution}
    \\
    HMPA (\genref{gen:hmpa})
    & \cmark &  &  &  &  
    & $O(\abs{V})$ & 
    Dynamic Full-Hypergraph & 
    \patternref{pat:community} 
    \\    
    \textsc{HyperFF} (\genref{gen:hyperff})
    & \cmark &  &  &  &  
    & $O(1)$ & 
    Dynamic Full-Hypergraph & 
    \ptref{pattern:heavy_tailed_deg_dist}, \ptref{pat:heavy_tailed_size_dist},
    \ptref{pat:heavy_tailed_intersection_size_distribution},
    \ptref{pat:singular_value_distribution},
    \ptref{pat:diminishing_overlap},
    \ptref{pat:densification},
    \ptref{pat:diameter}
    \\    
    \thera (\genref{gen:thera})
    & \cmark &  & \cmark &  &  
    & $O(1)$ & 
    Dynamic Full-Hypergraph &
    \ptref{pattern:heavy_tailed_deg_dist}, 
    \ptref{pat:heavy_tailed_size_dist},
    \ptref{pat:heavy_tailed_intersection_size_distribution},
    \patternref{pat:high_transitivity}
    \\    
    DARH (\genref{gen:darh})
    & \cmark &  &  &  &  
    & $O(\abs{V})$ &
    Dynamic Full-Hypergraph &
    \ptref{pat:temporal_locality}
    \\
    \midrule
    \midas (\genref{gen:midas})
    & & & & & \cmark
    & $O(1)$ &
    Static Sub-Hypergraph &
    \ptref{pattern:heavy_tailed_deg_dist}, 
    \ptref{pat:heavy_tailed_size_dist},
    \ptref{pat:heavy_tail_group_deg},
    \ptref{pat:heavy_tailed_intersection_size_distribution}, \ptref{pat:singular_value_distribution}
    
    \\
    HRW (\genref{gen:hrw})    
    & & & & & \cmark
    & $O(1)$ &
    Static Sub-Hypergraph &
    \ptref{pattern:heavy_tailed_deg_dist}, 
    \ptref{pat:heavy_tailed_size_dist} 
    \\
    \midrule
    TRHC (\genref{gen:trhc})
    & & & & & \cmark
    & $O(1)$ &
    Dynamic Sub-Hypergraph &
    \patternref{pat:egonet_temp_locality}, \patternref{pat:egonet_anthropic_principle},
    \patternref{pat:egonet_novelty_rate}
    \\
    CRU (\genref{gen:cru})
    & & & & & \cmark
    & $O(\abs{E})$ &
    Dynamic Sub-Hypergraph &
    \ptref{pat:heavy_tail_group_deg},
    \ptref{pat:repeat_behaviors},
    \patternref{pat:temporal_locality}
    \\
    \bottomrule
\end{tabular}
}
}
\vspace{0.5ex}

{\raggedright \footnotesize {*}If a generator has DD (or EH) in their input information, then patterns regarding node degrees (e.g., \ptref{pattern:heavy_tailed_deg_dist}) can be preserved in a hard-wired manner; similarly, we hyperedge-size-related patterns (e.g., \ptref{pat:heavy_tailed_size_dist}) can be preserved in a hard-wired manner by generators with SD (or EH) in their input information.\par}
 {\raggedright \footnotesize {**}It can be reduced to $O(1)$ by using the automatic hyperparameter selection method~\citep{lee2021hyperedges}.\par} 
\end{table}
\endgroup

\generator{\hyperlap.}\label{gen:hyperlap}
\hyperlap, proposed by~\cite{lee2021hyperedges}, is based on the overlapping patterns of hyperedges in real-world hypergraphs (see \patternref{pat:hyperedge_homogeneity} and \patternref{pat:density_overlapness}).
It can be seen as a multilevel extension of \hypercl (see \nmref{nullmodel:hypercl}), where the extension helps reproduce the overlapping patterns.

\begin{itemize}
    \item \textbf{Algorithm summary:}
    The nodes are organized in multiple levels, where each level contains all the nodes but in different granularities.
    {Specifically, from top to bottom, each group at one level is divided into two groups in the next level, and thus deeper (i.e., closer-to-bottom) levels contain more groups.}
    When generating each hyperedge, \hyperlap first samples a level and then chooses a group in the level.
    Nodes in the chosen group are sampled to fill the hyperedge, with probability proportional to the degrees.    
    \item \textbf{Intuitions:}
    {As hyperedges are generated based on groups,}
    nodes within the same group are structurally similar, which makes the generated hyperedges have high homogeneity (see \patternref{pat:hyperedge_homogeneity}).    
    {Since each ego-network tends to contain structurally similar nodes, which co-appear in many hyperedges,}
    the density (see \msref{measure:density}) and overlapness (see \msref{measure:overlapness}) of each ego-network (see \patternref{pat:density_overlapness}) naturally arise.
    Finally, pairs or triplets in a small group at a deep (i.e., close-to-bottom) {level} {also belong to the same groups at all the shallower (i.e., closer-to-top) levels}, and thus they
    are more frequently chosen to form hyperedges together.
    Therefore, more hyperedges contain and overlap at such pairs or triplets,
    which implies skewed heavy-tailed pair-of-nodes and triplet-of-nodes degree distributions (see \ptref{pat:heavy_tail_group_deg}).
    \item \textbf{Notes:}
\cite{lee2021hyperedges} also proposed \hyperlapp, which additionally includes an automatic hyperparameter selection scheme on top of \hyperlap.
\end{itemize}

\generator{CIGAM.}\label{gen:cigam}
CIGAM (continuous influencer-guided attachment model), proposed by~\cite{Papachristou2022CoreperipheryMF}, aims to capture core-periphery structures in real-world hypergraphs (see \patternref{pattern:core_periphery}).

\begin{itemize}
    \item\textbf{Algorithm summary:}
    Each node is assigned a prestige value, and
    each potential hyperedge is generated independently,
    where the sampling probability is determined by the prestige values of the constituent nodes.
    \item\textbf{Intuitions:}
    The nodes with high prestige values are supposed to be ``core nodes''.
    {Each generated hyperedge is likely to contain core nodes}, which implies a core-periphery structure (see \patternref{pattern:core_periphery}).
    Only the maximum prestige value in each potential hyperedge is considered, in order to reduce the computational complexity in likelihood estimation.
    \item\textbf{Discussions:}
    The idea of assigning prestige values to nodes might be combined with other hypergraph generators to capture more properties, such as motifs (see \ccref{concept:HO-motifs} and \ccref{concept:H-motifs}).
\end{itemize}

\generator{\textsc{Hyper}-$dK$.}\label{gen:hyper_dk}
The \textsc{Hyper}-$dK$ series, proposed by 
\cite{nakajima2021randomizing}, are a family of reference models for hypergraphs, which extends the $dK$ series~\citep{mahadevan2006systematic} for pairwise graphs.
They generate hypergraphs preserving the given local properties of nodes and hyperedges (e.g., \patternref{pattern:heavy_tailed_deg_dist} and \patternref{pat:heavy_tail_group_deg}).

\begin{itemize}
    \item\textbf{Algorithm summary:}
    They use two hyperparameters $d_v \in \setbr{0, 1, 2, 2.5}$ and $d_e \in \setbr{0, 1}$.
    A higher $d_v$ value means higher-order information regarding node degrees is preserved, while a higher $d_e$ value means higher-order information regarding hyperedge sizes is preserved.
    For example, 
    $d_v = 0$ (or $d_e = 0$)
    represents preserving the average degree (or the average hyperedge size),
    $d_v = 1$ (or $d_e = 1$) represents preserving individual node degrees (or individual hyperedge sizes),
    and $d_v = 2$ represents preserving joint degrees of node pairs.
    The cases with $d_v \in \setbr{0, 1}$ (at most preserving the degree distribution) are straightforward, and they are very similar to configuration hypergraph models~\citep{chodrow2020configuration} (see \nmref{nullmodel:config_models}).
    For $d_v \in \setbr{2, 2.5}$ (preserving some higher-order patterns), the generator starts with $d_v = 1$ and then uses edge rewiring to improve the corresponding higher-order pattern while maintaining the degree distribution.
    \item\textbf{Discussions:}
    The higher-order patterns are preserved %
    {by direct manipulation,} i.e., edge rewiring, which might be less efficient.
Proposing local mechanisms that preserve such higher-order patterns would be an interesting future direction.
\end{itemize}

\generator{AMHM.}\label{gen:assortative}
AMHM (assortative mixing hypergraph model), proposed by~\cite{landry2022hypergraph}, aims to control the extent of assortativity (see \measureref{measure:assortativity}).

\begin{itemize}
    \item\textbf{Algorithm summary:}
    On top of a naive configuration model (see \nullmodelref{nullmodel:config_models}) that only preserves node degrees and hyperedge sizes, AMHM further uses a function to control the edge probabilities among nodes with different combinations of degrees.
    \item\textbf{Intuitions:}
    Assortativity can be directly controlled by the function. For example, we can increase assortativity by increasing the edge probabilities among nodes with similar degrees.
    \item\textbf{Discussions:}
    As discussed by the authors, AMHM does not capture higher-order connection patterns beyond individual hyperedges, e.g., encapsulation (see \patternref{pat:encapsulation})
    and overlap (see \patternref{pat:density_overlapness}).
\end{itemize}

\generator{HSBM.}\label{gen:hyperSBM}
HSBM (hypergraph stochastic block model), proposed by~\cite{ghoshdastidar2017consistency}, can generate community structure (see \conceptref{concept:hypergraph_community} and \patternref{pat:community}) and is useful for community detection.
\begin{itemize}
    \item\textbf{Algorithm summary:}
    Nodes are divided into multiple groups (i.e., communities), and edges are generated with probabilities as a function of the membership combinations of nodes.
    \item\textbf{Intuitions:}
    The strength of community structure can be directly tuned by the function. For example, we can generate hypergraphs with strong community structure by increasing the edge probability for nodes in the same group.    
    \item\textbf{Discussions:}
    More generalized models, such as the hypergraph censored block model~\citep{ahn2019community}, the sub-hypergraph stochastic block model~\citep{liang2021information}, the hypergraph degree-corrected stochastic block model~\citep{chodrow2021generative}, and
    hypergraph simultaneous generators~\citep{pedrood2022hypergraph},
    have also been considered.
\end{itemize}
\color{black}

\generator{RNHM.}\label{gen:rnhm}
RNHM (random nested hypergraph model), proposed by~\cite{kim2023contagion}, aims to control the degree of hyperedge nestedness (i.e., encapsulation; see \ccref{concept:encaps_graph} and \patternref{pat:encapsulation}).

\begin{itemize}
    \item\textbf{Algorithm summary:}
    First, many large hyperedges are generated, together with all their subsets.
    Then, rewiring steps are conducted.
    In each rewiring step, RNHM
    (1) first randomly chooses a hyperedge $e$, and
    (2) chooses a node $v \in e$ and randomly replaces some nodes in $e$ with the same number of other nodes originally not connected (see \bgref{background:paths_connectivity}) with $v$.
    The probability of each hyperedge being chosen %
    and the number of rewiring steps are tunable.
    \item\textbf{Intuitions:}
    In the initial state, the nestedness (i.e., encapsulation; see \ccref{concept:encaps_graph}) is maximized.
    Each rewiring step breaks some nested substructures.
    Tuning allows us to control the degree of nestedness in the final generated hypergraph.
    \item\textbf{Discussions:}
    Although RNHM has been further discussed and utilized by~\cite{larock2023encapsulation},
    how to determine the parameters in RNHM to better preserve the observed patterns is still waiting to be covered.
\end{itemize}

\generator{HOC.}\label{gen:hoc}
HOC (higher-order-connected), proposed by~\cite{kim2022higher}, is based on the observations of higher-order connectivity in real-world hypergraphs (see \patternref{pat:higher_order_conn}).

\begin{itemize}
    \item\textbf{Algorithm summary:}
    The nodes are preassigned to subgroups of size two.
    HOC starts with a given number of empty hyperedges,
    and HOC repeats the following process until the average size of hyperedges reaches a given number:
    HOC first randomly chooses a hyperedge $e$, then
    with some probability $1 - p$, a random node is added into $e$, 
    and with the remaining probability $p$, all the nodes in a random subgroup are added into $e$.
    \item\textbf{Intuitions:}
    When the value of $p$ increases, the nodes in the same subgroup frequently co-exist in multiple hyperedges, which results in higher-order components where the intersection of multiple hyperedges has a large size (see \patternref{pat:higher_order_conn}).
    \item\textbf{Discussions:}
    Each node is assigned to a single subgroup, while the idea of %
    overlapping communities~\citep{contisciani2022inference} (see \ccref{concept:hypergraph_community}) 
    might be considered to improve the flexibility of this model.
\end{itemize}

\subsubsection{Dynamic generators}
We are now going to introduce \textit{dynamic full-hypergraph generators}.

\generator{\hyperpa.}\label{gen:hyperpa}
\hyperpa, proposed by~\cite{do2020structural}, is based on observations about $k$-level decomposed graphs (see \patternref{pat:heavy_tail_group_deg}, \patternref{pat:high_transitivity}, and \patternref{pat:singular_value_distribution}).
It is a group-wise extension of preferential attachment \citep{barabasi1999emergence}) where the key idea is that new nodes are more likely attached to existing high-degree nodes, making the ``rich'' nodes ``richer''.
For example, researchers who have co-authored many papers are likely to share common interests, which leads to more future collaborations.

\begin{itemize}
    \item\textbf{Algorithm summary:}
    For each node $v$, \hyperpa first samples the number of ``new'' hyperedges.
    For each ``new'' hyperedge, \hyperpa samples a hyperedge size $s$, and then attaches the node $v$ to a group of size $(s-1)$ with probability proportional to the group degrees {(see \msref{measure:group_deg})} using preferential attachment.
    \item\textbf{Intuitions:}
    Preferential attachment is known to be able to produce graphs with
    skewed degree distributions (see \ptref{pattern:heavy_tailed_deg_dist}),
    high clustering coefficients, %
    small diameters, %
    etc.
    Intuitively, \hyperpa also produces hypergraphs with similar patterns generalized to hypergraphs (see \patternref{pat:heavy_tail_group_deg}, \patternref{pat:high_transitivity}, and \patternref{pat:singular_value_distribution}).
    Notably, preferential attachment is done in a group-wise manner in \hyperpa, which produces
    skewed group (and also individual) degrees (see
    \ptref{pattern:heavy_tailed_deg_dist} and \patternref{pat:heavy_tail_group_deg}).
    \item\textbf{Discussions:}
    \hyperpa generates hyperedges in a dynamic way. 
    However, the timestamps of hyperedges are considered neither in observations nor in evaluations. 
    Extending the observations and evaluations to temporal hypergraphs would be an interesting future direction.
    {Refer to \citep{roh2023growing} for more general discussions on preferential attachment in hypergraphs.}
\end{itemize}

\generator{HMPA.}\label{gen:hmpa}
HMPA (high-modularity preferential attachment), proposed by~\cite{giroire2022preferential}, also uses the idea of preferential attachment.
The generator also considers community structures (see \ptref{pat:community}) with high modularity (see \msref{measure:modularity}), which are observed in real-world hypergraphs.

\begin{itemize}
    \item\textbf{Algorithmic summary:}
    The nodes are explicitly partitioned into communities (see \ccref{concept:hypergraph_community}).
    At each time step, either a new node is generated and attached to a community,
    or a new hyperedge is generated with existing nodes.
    When generating each new hyperedge, HMPA first samples a group of communities and then determines the number of nodes to be chosen from each community.
    Within each community, the nodes are chosen with probability proportional to their degrees, in a preferential-attachment way.
    \item\textbf{Intuitions:}
    Community structures are directly imposed by partitioning nodes.
    One can manipulate the sampling probabilities to encourage more hyperedges consisting of nodes in a small number of communities or even a single community, which implies high modularity.
    \item\textbf{Discussions:}    
    In this generator, the communities are disjoint and the community memberships are fixed. It might be beneficial to have more flexible community structures, such as overlapping communities~\citep{contisciani2022inference, cazabet2010detection} (see \ptref{pat:community}).
\end{itemize}

\generator{\hyperff.}\label{gen:hyperff}
\hyperff, proposed by~\cite{kook2020evolution}, is based on observations regarding the evolution of real-world hypergraphs (see \patternref{pat:diminishing_overlap}, \patternref{pat:densification}, and \ptref{pat:diameter}).
As the name indicates, this generator is inspired by the forest fire model~\citep{leskovec2007graph} on pairwise graphs.

\begin{itemize}
    \item\textbf{Algorithm summary:}
    At each time step, a new node joins, and an existing node is chosen as the ambassador node, {where a ``forest fire'' begins}. The forest fire is spread {stochastically} through existing hyperedges.
    When it terminates, %
    {from each burned node, a new forest fire begins, and the nodes burned in this round together form a hyperedge with the new node.}
    \item\textbf{Intuitions:}
    The authors were motivated by real-world scenarios in co-authorship networks.
    At each time step, the new node represents a new student joining a research community, the ambassador node represents the new student's supervisor,
    and the forest-fire-like process represents the real-world process where researchers collaborate with each other.
    \item \textbf{Notes:}
    {\cite{ko2022growth} also developed a simplified and mathematically tractable version of \hyperff, which leads to closed-form equations for expected hyperedge sizes, hyperedge numbers, and node degrees.}
    However, the simplified version has a weaker ability to reproduce real-world patterns, especially regarding shrinking diameters (see \ptref{pat:diameter}).
    \item \textbf{Discussions:}
    \hyperff focuses on preserving real-world hypergraph patterns in a macroscopic way.
    It would be an interesting future direction to further consider microscopic patterns, such as hyperedge ordering or repetition.    
\end{itemize}

\generator{\thera.}\label{gen:thera}
\thera (transitive hypergraph generator), proposed by~\cite{kim2023transitive}, is based on the observations regarding the transitivity of real-world hypergraphs (see \patternref{pat:high_transitivity}).

\begin{itemize}
    \item\textbf{Algorithm summary:}
    The nodes are organized in a hierarchical structure (specifically, a tree) with multiple levels,
    where the nodes are divided into disjoint levels with deeper (i.e., closer-to-leaf) levels containing more nodes,
    {and the nodes at each level} are split into disjoint groups.
    The group size is the same across different levels, and thus there are more groups at deeper levels.
    When generating a hyperedge, 
    with some probability, \thera generates it locally within a group, and
    with the remaining probability, \thera generates it globally {within the entire node set}, with the nodes in shallower levels more likely to be chosen. 
    
    \item\textbf{Intuitions:}
    The community structure (see \ccref{pat:community}) naturally gives high transitivity (see \ptref{pat:high_transitivity}). %
    The hierarchical structure allows nodes at different levels to be chosen with different probabilities, which implies realistic skewed degree distributions, specifically a large number of small-degree nodes and a small number of large-degree nodes  (see \ptref{pattern:heavy_tailed_deg_dist}).
    \item\textbf{Discussions:}
    The hyperparameters of \thera {have to be} chosen manually. It would be desirable to have a fitting algorithm that automatically chooses hyperparameters. %
\end{itemize}

\generator{DARH.}\label{gen:darh}
DARH (discrete auto-regressive hypergraph), proposed by~\cite{gallo2023higher}, is based on the temporal locality in real-world hypergraphs (see \ptref{pat:temporal_locality}).

\begin{itemize}
    \item\textbf{Algorithm summary:}
    In DARH, at each time step $t$,
    for each potential hyperedge $e$:
    (1) with some probability $q$,
    a random previous time step $t' < t$ is sampled and $e$ exists at $t$ if and only if $e$ existed at $t'$;
    and
    (2) with the remaining probability $1 - q$,
    we do another Bernoulli trial with success probability $y$, and $e$ exists if and only if the Bernoulli trial succeeds.   
    The values of $q$ and $y$ are the same for hyperedges of the same size, while the values can be different for hyperedges of different sizes.
    \item\textbf{Intuitions:}
    The temporal dependencies (see \ptref{pat:temporal_locality}) are directly imposed by the case where DARH copies the states from previous time steps.    
    Hyperedges of the same size are supposed to show similar temporal patterns since they have the same parameters $q$ and $y$.
    \item\textbf{Discussions:}
    They also generalized DARH to further incorporate temporal dependencies across different hyperedge sizes (see \ptref{pat:temporal_locality}).
\end{itemize}

\subsection{Sub-hypergraph generators}\label{sec:generators:sub}
Now, we introduce \textit{sub-hypergraph generators}.
Unlike full-hypergraph generators, sub-hypergraph generators output a subgraph of a given hypergraph while preserving some properties.

\subsubsection{Static generators}
We shall introduce \textit{static sub-hypergraph generators} below.

\generator{\midas.}\label{gen:midas}
\midas (minimum degree biased sampling of hyperedges), proposed by~\cite{choe2022midas,choe2024representative}, aims to generate a representative subhypergraph of a given hyperedge, 
where the properties  {(e.g., \ptref{pattern:heavy_tailed_deg_dist}, 
    \ptref{pat:heavy_tailed_size_dist},
    \ptref{pat:heavy_tail_group_deg},
    \ptref{pat:heavy_tailed_intersection_size_distribution}, and \ptref{pat:singular_value_distribution})} of the given hypergraph are well-preserved. %

\begin{itemize}
    \item\textbf{Algorithm summary:}
    In \midas, hyperedges are sampled one by one.
    The sampling probability of each hyperedge is determined by the minimum node degree in it.
    A trained linear regression model is used to automatically tune the extent of bias towards high-degree nodes.
    \item\textbf{Intuitions:}
    \midas extends random hyperedge sampling by introducing bias w.r.t node degrees.
    The design of \midas is motivated by two observations:
    (1) random hyperedge sampling works well overall but fails to generate subhypergraphs with high connectivity (see \bgref{background:paths_connectivity}) and enough high-degree nodes (see \ptref{pattern:heavy_tailed_deg_dist}), and
    (2) {preserving the degree distribution is closely associated with preserving several other hypergraph properties.}
    \item\textbf{Notes:}
    Although \midas only directly considers bias w.r.t node degrees (see \ptref{pattern:heavy_tailed_deg_dist}), it preserves many other hypergraph properties {(e.g.,
    \ptref{pat:heavy_tailed_size_dist},
    \ptref{pat:heavy_tail_group_deg},
    \ptref{pat:heavy_tailed_intersection_size_distribution}, and \ptref{pat:singular_value_distribution})}.
    \item \textbf{Discussions:}
    It would be interesting to study the deeper reasons behind {the strong connection between preserving node degrees and preserving other hypergraph properties.}
    Besides, considering temporal information (if available) might be beneficial to the performance.
\end{itemize}

\generator{HRW.}\label{gen:hrw}
HRW (hybrid random walk), proposed by~\cite{zhang2023efficiently}, uses sampling via random walks on hypergraphs.
The sampled (i.e., visited) nodes and hyperedges are used to estimate the statistics, such as node-degree and hyperedge-size distributions (see \bgref{background:degree_size_distribution}) of an input hypergraph.

\begin{itemize}
    \item\textbf{Algorithm summary:}    
    HRW obtains both node and hyperedge samples by Markov chain Monte Carlo (MCMC) with separate node and hyperedge transitions.
    The statistics of an input hypergraph are estimated based on the sampled nodes and hyperedges.
    \item\textbf{Intuition:}    
    Naive construction of a Markov chain on hyperedges has high time and space complexity.
    HRW separates node and hyperedge transitions to avoid considering combinations of nodes and hyperedges and thus reduces the complexity of the state space.  
    \item\textbf{Notes:}
    They also proposed techniques to improve the sampling efficiency and estimation accuracy of HRW.
    The techniques include a non-backtracking strategy to speed up the convergence using lifted Markov chains~\citep{chen1999lifting}
    and a skipping strategy to speed up the transitions. %
\end{itemize}

\subsubsection{Dynamic generators}
We are now going to introduce \textit{dynamic sub-hypergraph generators}.

\generator{TRHC.}\label{gen:trhc}
TRHC (temporal reconstruction hill climbing), proposed by \cite{comrie2021hypergraph}, is mainly based on observations regarding hypergraph ego-networks (see \patternref{pat:egonet_temp_locality}, \patternref{pat:egonet_anthropic_principle}, and \patternref{pat:egonet_novelty_rate}).

\begin{itemize}
    \item\textbf{Algorithm summary:}
    Given an ego-network, TRHC assigns a temporal order to the hyperedges in the given ego-network.
    Starting from an initial order, TRHC keeps swapping hyperedge pairs to improve the ``fitness'' of different orders,
    where the ``fitness'' is evaluated by a supervised model.
    The process terminates when no improvement is possible.
    \item\textbf{Intuitions:}
    The supervised model %
    is trained so that it is supposed to give higher ``fitness'' values to temporal orders matching the observations better.
    Therefore, improving the ``fitness'' is supposed to make the temporal order more similar to the observed patterns.
    \item\textbf{Discussions:}
    This generator is limited to predicting orders of a fully given group of hyperedges, and it is also limited to hypergraph ego-networks.
    Besides, in real-world scenarios, 
    multiple different orders of a group of hyperedges might be equally likely and realistic.    
    Studying partial orders (e.g., causal relations~\citep{ma2022learning}) might be an interesting direction.
\end{itemize}

\generator{CRU.}\label{gen:cru}
CRU (correlated repeated unions), proposed by~\cite{benson2018sequences}, is mainly based on observations regarding temporal behavior in real-world temporal hypergraphs (see \patternref{pat:temporal_locality} and \patternref{pat:simplicial_closure}).

\begin{itemize}
    \item\textbf{Algorithm summary:}
    CRU generates hyperedges {in a subhypergraph (e.g., star ego-network; see \ccref{concept:ego_networks})} in a sequential manner.
    For each new hyperedge, its size and the new nodes in it that have never appeared before are assumed to be known.
    To find the remaining nodes to fill the hyperedge, CRU samples {a hyperedge} from the existing ones, where more recent hyperedges are more likely to be chosen.    
    {Each node in the selected hyperedge is independently copied into the new hyperedge with a sampling probability of $p$ until the new hyperedge is full.
    If the new hyperedge is not filled, the same process is repeated.
    }
    \item\textbf{Intuitions:}
    Sampling from existing hyperedges establishes the temporal correlation (see \patternref{pat:temporal_locality}) between new hyperedges and existing hyperedges. {The extent of temporal locality is controlled by the bias towards recent hyperedges, and the extent of correlation (i.e, repetition of subsets) is determined by the sampling probability $p$.}
    \item\textbf{Discussions:}
    The required input information includes the size of each new hyperedge and the number of new nodes in it, which might be unrealistically strong.
    It would be a challenging yet interesting future direction to preserve the considered patterns with a weaker oracle. %
\end{itemize}

\section{Future applications and directions}\label{sec:future}

In this section, we discuss future applications and directions of hypergraph mining, especially hypergraph patterns.
We mainly discuss existing applications and research topics related to graph mining, especially the graph counterparts of what we have discussed in this survey.
Since most hypergraph patterns are generalized from graph patterns, we expect that many existing applications and directions of graph mining will also be extended and generalized to hypergraphs in the future.
For a more in-depth discussion with additional references, refer to the supplementary document~\citep{supplementary}.

\subsection{Applications to algorithmic design}\label{sec:future:algo_design}
Graph mining patterns have inspired innovative graph algorithms for real-world applications, which we expect can extend to hypergraph mining.

\smallsection{Degree distributions.}
The observation that real-world graphs usually {exhibit heavy-tailed degree distributions}
has been used for the design of graph algorithms,
including distributed graph algorithms~\citep{salihoglu2013gps},
graph traversal algorithms~\citep{zhang2018degree}, etc.
Based on the similar pattern in hypergraphs (see \ptref{pattern:heavy_tailed_deg_dist} and \ptref{pat:singular_value_distribution}), the above applications are possibly extendable to hypergraphs.

\smallsection{Temporal locality.}
In many real-world temporal graphs, the temporal locality is observed, where
edges appearing within a smaller temporal window are more likely to interact.
This property has been used for {designing efficient algorithms for} 
triangle counting~\citep{lee2020temporal}
and graph traversal~\citep{koohi2021exploiting}.
Several patterns related to temporal locality have been observed in hypergraphs {(see \ptref{pat:temporal_locality} and \ptref{pat:egonet_temp_locality}),} %
and we expect such patterns to be useful in the design of algorithms for temporal hypergraphs.

\smallsection{Diameters.}
Small diameters in real-world graphs have been considered in designing algorithms for large-scale graph mining~\citep{kang2011pegasus}.
Therefore, shrinking diameters observed in real-world hypergraphs {(see \ptref{pat:diameter})} are also possibly useful for large-scale hypergraph mining.

\subsection{Applications to machine learning}\label{sec:future:ml}
Graph patterns have also been widely used in machine learning on graphs, suggesting the potential usefulness of hypergraph patterns in machine learning on hypergraphs.

\smallsection{Graph neural networks and general feature representation.}
One of the most common topics in machine learning on graphs is feature representation, where graph neural networks (GNNs) are often used.
Many graph properties and patterns have been considered for enhancing the performance of GNNs,
including
degree distributions~\citep{liu2019hyperbolic} and
assortativity~\citep{suresh2021breaking}.
We expect hypergraph patterns to be useful, like their graph counterparts, in (hyper)graph neural networks.
and general feature representation in hypergraphs. %

\smallsection{Link prediction and community detection.}
Link prediction %
and community detection %
are two traditional machine-learning problems on graphs.
Many graph patterns have been used in those two problems, including
assortativity~\citep{ciglan2013community} and
graph motifs~\citep{rotabi2017detecting}.
We look forward to seeing hypergraph patterns be used for these two tasks.

\smallsection{Anomaly detection.}
Anomaly detection %
is another traditional machine-learning problem, and
graph-based anomaly detection~\citep{akoglu2015graph} is a popular subtopic.
Many graph patterns have been used in graph-based anomaly detection, including
graph motifs~\citep{noble2003graph} and
$k$-cores~\citep{shin2016corescope}.
We anticipate more usage of hypergraph patterns for this application.

\smallsection{Recommendation.}
Graphs are an important tool for building recommender systems, a long-standing research topic in machine learning, where
many graph patterns, including
graph motifs~\citep{gupta2014real}
and
the structure of ego-networks~\citep{epasto2015ego}, 
have been used.
Hypergraphs are also useful for this task, especially 
bundle recommendation~\citep{zhu2014bundle}.
We await more applications of hypergraph patterns for recommender systems.

\subsection{Analysis and mining of generalized hypergraphs}\label{sec:future:gen_hgs}
In this survey, we have mainly discussed simple hypergraphs (i.e., undirected and unweighted ones).
Below, we would like to discuss several types of generalized hypergraphs.

\smallsection{Directed hypergraphs.}
Directed hypergraphs,
{where nodes within each hyperedge are partitioned into a source set and a destination set},
have been studied in the fields of mathematics and theoretical computer science. %
Recently, \cite{kim2022reciprocity} extended the concept of reciprocity to directed hypergraphs and studied related patterns within real-world directed hypergraphs.
We expect more patterns to be explored on directed hypergraphs.

\smallsection{Weighted hypergraphs.}
Most works mentioned in this survey deal with unweighted real-world hypergraphs or explicitly preprocess the datasets into unweighted ones, although some take the repetition of hyperedges into consideration~\citep{benson2018sequences,lee2021thyme+}.
Recently, there also has been a growing interest in hypergraphs with edge-dependent vertex weights (where a node can have different weights in different hyperedges)~\citep{Chitra2019RandomWO}.
We expect more patterns to be explored on weighted hypergraphs.

\section{Conclusions}\label{sec:conclusion}
Hypergraphs are a valuable mathematical framework for modeling complex group interactions in a variety of real-world scenarios. The inherent complexity of hypergraphs brings both opportunities and challenges for hypergraph mining, which has recently attracted increasing attention. This survey thoroughly examines the advancements so far in hypergraph mining and provides a comprehensive overview of the patterns, tools, and generators for hypergraphs. We also present complete taxonomies to improve the structured understanding of each aspect. Finally, we suggest several research directions. We hope that this overview will provide researchers and practitioners with valuable resources and insights that will advance both fundamental research and practical applications of hypergraphs across various fields.

\section*{Acknowledgements}
This work was supported by National Research Foundation of Korea (NRF) grant funded by the Korea government (MSIT) (No. NRF-2020R1C1C1008296) and Institute of Information \& Communications Technology Planning \& Evaluation (IITP) grant funded by the Korea government (MSIT)
(No. RS-2024-00438638, EntireDB2AI: Foundations and Software for Comprehensive Deep Representation Learning and Prediction on Entire Relational Databases)
(No. RS-2019-II190075, Artificial Intelligence Graduate School Program (KAIST)).

\bibliographystyle{ACM-Reference-Format}
\bibliography{ref}

\end{document}